\documentclass[amsmath,amssymb,aps,prx,reprint,showpacs,floats,superscriptaddress,longbibliography]{revtex4-2}
    \usepackage[utf8]{inputenc}
    \usepackage[table,xcdraw]{xcolor}
    \usepackage{graphicx}
    \usepackage{amsmath, amsthm, amssymb, amsfonts}
    \usepackage{mathtools}
    \usepackage[colorlinks=true,citecolor=blue,linkcolor=blue,urlcolor=blue]{hyperref}
    \usepackage{footnote}
    \usepackage{xifthen}    
    \usepackage{capt-of}
    \usepackage{subfigure}
    \usepackage{physics}
    \usepackage{adjustbox}
    \usepackage{tikz}
    \pdfoutput=1 
    \usetikzlibrary{external}
    \usetikzlibrary{arrows.meta}
    \usetikzlibrary{decorations.pathmorphing}
    \usetikzlibrary{positioning}
    \usetikzlibrary{backgrounds}
    \usetikzlibrary{calc}
    \usetikzlibrary{shapes.symbols}
    \usetikzlibrary{shapes.arrows}
    \usetikzlibrary{fadings}

    \newcommand{\cbqty}[1]{\left\{#1\right\}}

    \newcommand{\sn}[1]{\mathfrak{#1}}
    \newcommand{\opr}[1]{#1}
    \newcommand{\sigz}{\opr{\sigma}^z}
    \newcommand{\sigx}{\opr{\sigma}^x}
    
    \newcommand{\HH}{\opr{H}}
    \newcommand{\VV}{\opr{V}}
    \newcommand{\bbi}{\opr{\mathbb{I}}}
    \newcommand{\Sop}[1]{#1\otimes\bbi_\sn{E}}
    \newcommand{\Eop}[1]{\bbi_\sn{S}\otimes#1}

\begin{document}
\title{Classifying two-body Hamiltonians for Quantum Darwinism}
\author{Emery~Doucet}
\email{emery.doucet@umbc.edu}
\affiliation{Department of Physics, University of Maryland, Baltimore County, Baltimore, MD 21250, USA}
\author{Sebastian~Deffner}
\email{deffner@umbc.edu}
\affiliation{Department of Physics, University of Maryland, Baltimore County, Baltimore, MD 21250, USA}
\affiliation{National Quantum Laboratory, College Park, MD 20740, USA}
\date{\today}
\begin{abstract}
Quantum Darwinism is a paradigm to understand how classically objective reality emerges from within a fundamentally quantum universe. Despite the growing attention that this field of research has been enjoying, it is currently not known what specific properties a given Hamiltonian describing a generic quantum system must have to allow the emergence of classicality. Therefore, in the present work, we consider a broadly applicable generic model of an arbitrary finite-dimensional system interacting with an environment formed from an arbitrary collection of finite-dimensional degrees of freedom via an unspecified, potentially time-dependent Hamiltonian containing at most two-body interaction terms. We show that such models support quantum Darwinism if the set of operators acting on the system which enter the Hamiltonian satisfy a set of commutation relations with a pointer observable and with one other. We demonstrate our results by analyzing a wide range of example systems: a qutrit interacting with a qubit environment, a qubit-qubit model with interactions alternating in time, and a series of collision models including a minimal model of a quantum Maxwell demon. 
\end{abstract}
\maketitle
%
%
\section{Introduction}
\label{sec:Intro}
%

At the fundamental level our universe is described by quantum physics, described by rules which seem to be profoundly at variance from the rules of the classical reality governing the everyday world. 
In the classical realm, measurements are objective and repeatable in the sense that many observers who each independently measure the same quantity of a classical system agree on their results.
The mechanism by which this objectivity emerges from the underlying rules of quantum mechanics can be understood through a framework initially developed by Zurek and collaborators, and which now has grown into a veritable field of modern research: quantum Darwinism \cite{Zurek03,Zurek09,BlumeKohout05,BlumeKohout06,Ollivier04,Ollivier05,Riedel2010PRL,Riedel2011NJP,Brandao2015NC,Ciampini2018PRA,Knott2018PRL,Milazzo2019PRA,Colafranceschi2020JPA,Touil22b,Lorenzo2020PRR,Girolami2022PRL}, which builds on the theory of decoherence \cite{Zurek2003RMP,Schlosshauer2005RMP,BreuerPetruccione} in open quantum systems to provide a description of emergent objectivity in quantum systems. See also a recent special issue summarizing the state of the art \cite{Zurek22} and featuring recent breakthroughs in the field \cite{QDAndFriends}.

Central to quantum Darwinism is the realization that no observer ever directly measures any system of interest $\sn{S}$ -- instead, any measurement uses the environment $\sn{E}$ in which the system is immersed as a channel to \emph{indirectly} probe the system.
Further, the observer usually cannot access the entire environment and instead infers the result of their measurement on $\sn{S}$ from some small environment fragment $\sn{F}$. 

\begin{figure}
    \centering
    \includegraphics[width=\columnwidth]{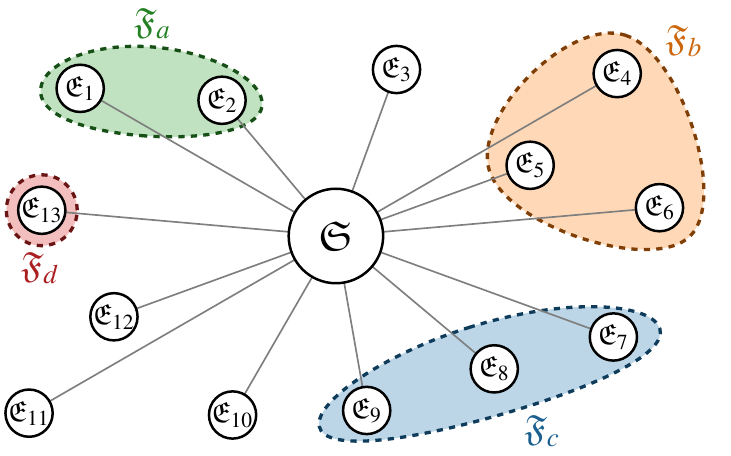}
    \caption{
    Schematic representation of the paradigm of quantum Darwinism. A system $\sn{S}$ is immersed in an environment $\sn{E}$, comprised of many subsystems $\{\sn{E}_1, \sn{E_2}, \dots\}$. Observers obtain information about the system \emph{indirectly} through measurements of fragments of the environment $\sn{F}$. Quantum Darwinism recognizes that the system-environment interaction leads to information about the system being redundantly encoded in the environment such that many observers can recover the classically accessible information about the system through measurements of distinct fragments of the environment (e.g., $\sn{F}_{a,b,c,d}$). Since this process allows many observers to reconstruct the same information with no back-action on the system, that information has become classical and objective.
    }
    \label{fig:DarwinismCartoon}
\end{figure}

This leads to a natural understanding of the emergence of classicality in a quantum model of a system and its environment: if the relevant information about the system necessary for a measurement is accessible to many observers who each capture independent small fragments of the environment, then the observers agree on their inferred measurement results and so that result is objective.
 
A schematic representation of this is depicted in Fig.~\ref{fig:DarwinismCartoon}. 
There is a central system $\sn{S}$ which is immersed in and interacting with an environment $\sn{E}$ which is itself comprised of a collection of environment degrees of freedom or sub-environments $\sn{E}_i$. 
Quantum Darwinism recognizes that this environmental interaction decoheres the system and redunantly encodes information about the decohered state in the environment.
Then, because of the redundant encoding, it is possible to reconstruct the accessible information about the decohered system state through measurements of small fragments of the environment $\sn{F}$.
Different observers who measure disjoint fragments of the environment (e.g., observer $a$ measures $\sn{F}_a$, observer $b$ measures $\sn{F}_b$ and so on) reconstruct the same information, and so that information is objective. 
 
\begin{figure}
    \centering
    \includegraphics[width=0.9\columnwidth]{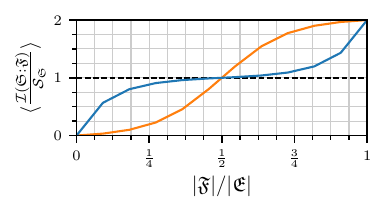}
    \caption{Plot of the mutual information between a subset $\sn{F}$ of an environment $\sn{E}$ (here 11 qubits) and a system $\sn{S}$ (a qubit), normalized by the entropy of the reduced system state and averaged over all fragments of that size. The ``classical plateau'' refers to the flattening of the blue curve about $|\sn{F}|/|\sn{E}| = 1/2$, indicating that the system-environment state allows classical objectivity, whereas the orange curve shows a more typical state which does not. Note that the small size of the simulated environment smooths out the curves; with a sufficiently large environment, the blue curve would be almost perfectly flat at one over almost the entire range of fragment sizes and the orange curve would jump from near zero to near two as the fragment size crosses half the environment. The data for this plot are for the examples E and G discussed in Sec.~\ref{ssec:AlternatingQubit}.}
    \label{fig:DarwinismPlateau}
\end{figure}

A convenient visualization of quantum Darwinism is found by plotting the average mutual information between environment fragments and the system as a function of the fragment size.
Figure~\ref{fig:DarwinismPlateau} presents such a plot for a model consisting of a qubit system interacting with an environment consisting of 11 qubits (this model will be studied in detail in Sec.~\ref{ssec:AlternatingQubit}).
For comparison we have plotted the normalized mutual information for a scenario which supports quantum Darwinism alongside a scenario that does not. 
In the Darwinistic case, we see the characteristic ``classical plateau'' emerging: once the environment fragment reaches a certain size the mutual information is very close to the system entropy, indicating that all the accessible information about the system is recoverable from fragments of that size.
This remains true for almost all fragment sizes until the observer measures essentially the entire environment, in which case the quantum correlations can be reconstructed and the mutual information rises. 
By contrast, for the case which does not support quantum Darwinism the mutual information curve is more of a step (though highly smoothed by the small environment size). 
This indicates that the system-environment state does not have any redundancy -- half the environment must be measured to recover the accessible information, therefore the encoding is non-redundant and there is no objectivity in this case.
 
In a system-environment model exhibiting quantum Darwinism, the information about the system that becomes objective corresponds to its projections onto the so-called \emph{pointer states} \cite{Zurek81,Zurek82,Zurek03,Zurek22}, defined in terms of a Hermitian pointer observable.
These are the states which are stable under the interaction of the system with its environment and which specify the basis in which the initial system state decoheres.
For example, consider a qubit which dephases in the $z$ basis through interaction with its environment. 
The system states stationary under the dynamics, i.e. the pointer states, are the eigenstates of the pointer operator $\sigz$.
 
Beyond simply decohering the system, objectivity requires 
that the joint system-environment state be a of \emph{singly-branching state} \cite{BlumeKohout05,BlumeKohout06,Modi12,Zwolak13,Korbicz21,Zurek22}. 
These are the only states for which the quantum discord \cite{Zurek01,Henderson01}, defined as the difference between two classically-equivalent expressions for the mutual information, between the system and environment fragments vanishes \cite{Touil22}.
Non-zero quantum discord indicates the presence of non-classical correlations, and in particular indicates that not all of the mutual information between subsystems (here the system and environment fragment) is accessible through measurements on one subsystem alone (e.g., through measurements of environment fragments).
The term \emph{strong quantum Darwinism} \cite{Korbicz21,Le19,Cakmak2021entropy} is sometimes used to refer to quantum Darwinism with the requirement that the system-fragment mutual information be accessible, i.e., that the system-environment state is of singly-branching form. 
Another somewhat more formal approach to understanding the emergence of objectivity in system-environment models relies on spectrum broadcast structures \cite{Horodecki15,Korbicz17,Korbicz21}, built from studying the measurement properties necessary for objectivity (in particular, non-disturbance). The two approaches ultimately yield the same conclusions \cite{Le19,Korbicz21}, and the results presented in this work are equally applicable to both.
 
To summarize, quantum Darwinism explains the emergence of classical objectivity from a quantum system interacting with an environment. 
The question of whether or not any given model of a system-environment model supports quantum Darwinism and so exhibits the emergence of classical objectivity is equivalent to understanding if the model supports a pointer basis and if the dynamics produce states of singly-branching form.
Though these requirements are fairly concrete, it is not currently known how they translate to detailed constraints on generic system-environment models which exhibit quantum Darwinism. 

Therefore, a classification of arbitrary models of a system interacting with some environment according to whether or not they support quantum Darwinism appears highly desirable. In fact, such a classification would be a very useful tool to help guide analyses of quantum-to-classical transitions and the boundary between the quantum and classical realm across a wide range of scenarios, for example in connecting quantum thermodynamic models with classical results.

\begin{figure}
    \centering
    \includegraphics[width=0.8\columnwidth]{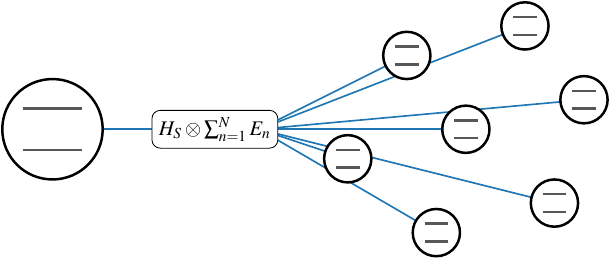}
    \includegraphics[width=0.8\columnwidth]{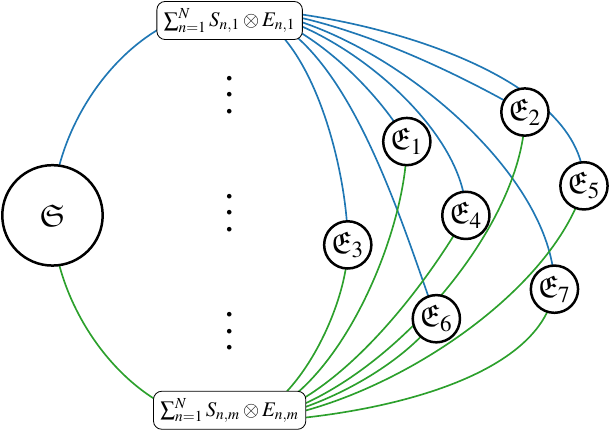}
    \caption{
    (Top) Schematic of the separable interaction structure necessary for a model of a qubit system interacting with an collection of qubits as an environment with two-body interactions to support Quantum Darwinism.
    (Bottom) Schematic of a generic $\mathfrak{S}\mathfrak{E}$ model demonstrating how the system may interact non-separably with the environment.
    }
    \label{fig:Schematic}
\end{figure}

A significant first step towards such a classification was taken in Ref.~\cite{Duruisseau23}, where the conditions necessary for objectivity to emerge in a qubit model with arbitrary two-body interactions were derived.
In the present work, we take a substantial step beyond the existing results in the literature and present a classification of generic system-environment models interacting with potentially time-dependent Hamiltonians -- subject only to the requirements that each subsystem is finite-dimensional and that the interaction Hamiltonian include only two-body interactions -- according to their ability to support quantum Darwinism. 

We emphasize that the class of Hamiltonians is sufficiently broad that the criteria we derive can be applied immediately to a large variety of existing models of both practical and theoretical importance, for example in quantum optics or quantum thermodynamics.
By presenting such a broad and comprehensive classification, we build a clear path forward to understand quantum Darwinism and the boundary between quantum and classical behavior across a wide range of relevant models. 
More importantly, the set of rules that must be satisfied if a model is to exhibit quantum Darwinism can be interpreted as guidelines when constructing new models to study the distinctions between quantum and classical dynamics, for example explicit quantum thermodynamics models such as quantum Maxwell's demons.
In this way,  our classification can be used not just as a tool to understand existing Hamiltonians but also as a recipe for future studies.
 
The outline of this work is as follows.
We begin by summarizing the existing classification of qubit models as presented in Ref.~\cite{Duruisseau23} in Sec.~\ref{sec:ExistingWork}, both to provide context for our results and as a point of comparison.
In Sec.~\ref{sec:Model} we consider time-independent system-environment Hamiltonians describing interactions between arbitrary finite dimensional subsystems, where we find the structural requirements placed on these models such that they support both a pointer basis and singly-branching states. 
We present results for a qutrit system coupled to an environment of qubits with several different interaction Hamiltonians, explicitly showing that objectivity can emerge with these more general interaction structures. 
We then extend our discussion to include arbitrary time-dependent Hamiltonians in Sec.~\ref{sec:TimeDep}.
The criteria for quantum Darwinism lead to constraints on the structure of these models which are straightforward generalizations of those which must be satisfied by time-independent models, which we demonstrate with simulations of several qubit models with two distinct interaction Hamiltonians which alternate in time.
Finally, in Sec.~\ref{sec:RepeatedContactModels} we consider collision models \cite{Strasberg17}, an important class of time-dependent models which can capture the behavior of a wide range of important quantum and thermodynamic systems, for example as a model for a system interacting with thermodynamic reservoirs of different types, or as a microscopic model for quantum measurement or open quantum dynamics.

%
\section{Quantum Darwinism in qubit models}
\label{sec:ExistingWork}
%
We begin by briefly summarizing Ref.~\cite{Duruisseau23}, to provide a basis for understanding how constraints are extracted from Hamiltonians and to indicate where and how the general case   differs from the qubit case.
 
Explicitly, the model considered in Ref.~\cite{Duruisseau23} consists of a single system qubit $\sn{S}$ interacting with an collection of $N$ qubits acting as an environment $\sn{E}$ which is then accessible to observers for measurement. 
The joint evolution of the system and environment is described by the Hamiltonian
\begin{align}
    \HH = 
        \sum_{i} \sum_{\alpha} B_i^\alpha\opr{\sigma}^\alpha_i
        + \sum_{i,j}\sum_{\alpha,\beta} J^{\alpha\beta}_{ij}\opr{\sigma}^\alpha_i\otimes\opr{\sigma}^\beta_j
    ,
\end{align}
where the subscripts $i,j$ correspond to qubit indices (with the system qubit being index $0$ and the environment qubits indices $1,\dots,N$) and the superscipts $\alpha,\beta \in \cbqty{x,y,z}$ select the different Pauli operators.
The coefficients $J^{\alpha\beta}_{ij}$ and $B_i^\alpha$ are assumed to be real values drawn from some fixed random distributions.
 
It was shown in Ref.~\cite{Duruisseau23} that there are three important requirements which must all be satisfied by Hamiltonians of this form for the qubit model to exhibit quantum Darwinism.
Firstly, for the Hamiltonian to support a pointer basis take the form of a ``parallel decoherent interaction'',
\begin{align}
    \HH = \Sop{\HH_S} + \Eop{\HH_E} + \HH_S \otimes \sum_{i=1}^N\opr{E}_i
    ,
    \label{eqn:QBModelH}
\end{align}
that is, the interaction between the system and the environment must be separable such that the operator acting on the system is proportional to the system Hamiltonian $\HH_S$. 
This is due to the fact that the system is a qubit -- once $\HH_S$ is fixed, the only possible pointer basis is its eigenbasis and the only operator which preserves that basis is $\HH_S$.
We will show that this structure is essentially unique to the case where the system is a qubit, and that with a general Hamiltonian there is not necessarily a relationship between separability of the interaction Hamiltonian and quantum Darwinism.
Schematically, Fig.~\ref{fig:Schematic}(a) shows a qubit model subject to a parallel decoherent interaction while Fig.~\ref{fig:Schematic}(b) shows the general structure we derive in this work for the interaction Hamiltonians compatible with quantum Darwinism in systems larger than a qubit.
 
Secondly, the vanishing of quantum discord between the system and environment fragments necessary for quantum Darwinism is only possible if the joint system-environment state is of singly-branching form \cite{Touil22}.
It should be noted that while this form of joint state is necessary and sufficient for the vanishing of quantum discord which is itself sufficient for objectivity to emerge in the sense of quantum Darwinism, it is not necessary.
For example, if the quantum discord is small but finite, the system-environment state is close in the Fubini-Study metric to a state of singly-branching form and  supports the emergence of objectivity.
In this work, we follow the example of Ref.~\cite{Duruisseau23} and search for Hamiltonians which support the singly-branching form (and so vanishing of discord) exactly for clarity and to simplify our derivations.
It should not be difficult to generalize our results to the case where this target is weakened, or to re-frame our arguments in terms of alternative perspectives on the emergence of objectivity, e.g. in terms of spectrum broadcast structures \cite{Horodecki15,Korbicz17,Korbicz21}. 
 
To satisfy this requirement, it is necessary that there be no interactions between any two environment qubits so that the information about the system which is encoded in the environment does not spread and become non-local. 
In the qubit model previously discussed, this requirement translates to a simple statement on the structure of the environment Hamiltonian alone. 
With a general Hamiltonian, the satisfaction of this requirement becomes a deeper statement due to the possibility that the system mediates some effective interactions between environment subsystems, especially when the interaction Hamiltonian is time-dependent. 
 
An intuitive understanding of this statement comes from revisiting the meaning of objectivity in quantum Darwinism given in the preceding paragraphs accompanying Fig.~\ref{fig:DarwinismCartoon}. 
The environment $\sn{E}$ is decomposed into the tensor product of a set of degrees of freedom, and the fragments which observers use to obtain information about the system have support on subsets of these degrees of freedom. 
Objectivity arises because the information about the system is redundantly encoded into the environment in such a way that it may be recovered from many different fragments. 
The primary issue that arises if intra-environmental interactions are introduced into the dynamics is that they cause the information within environment fragments to spread.
The details of how this information spreads can be understood through the lens of quantum information scrambling and entanglement growth \cite{Swingle18,Touil20,Duruisseau23,Touil24}, though for our purposes a detailed understanding is unnecessary.
What is important is that the initially local information contained in a fragment $\sn{F}$ that allowed an observer to reconstruct the system information becomes delocalized over time, which means it ceases to be possible to recover the system information from $\sn{F}$ \cite{Zurek14}.
Since it is no longer possible to determine the system state from small fragments, there can be no redundant encoding of information in the environment and hence no objectivity. 
 
Thirdly, the probability distributions from which the non-zero system-environment coupling strengths $J^{\alpha\beta}_{0i}$ are drawn from must have continuous support to ensure that the resulting dynamics describe an irreversible transfer of information from the system to the environment degrees of freedom. 
This remains true in the general case, with some caveats when the interaction Hamiltonian is allowed to be time-dependent.

\section{Generic time-independent model and constraints}
\label{sec:Model}
%

In this work, we consider a model of some system $\sn{S}$ which interacts with an environment consisting of $|\sn{E}| = N$ degrees of freedom.
We will restrict the system to be finite-dimensional, however we will not make any further assumptions about its size, internal structure or lack thereof, etc. 
Similarly, we do not make assumptions about the nature of each environment degree of freedom $\sn{E}_j$ beyond requiring that they also be finite-dimensional.
Importantly, we do not require that the different degrees of freedom in the environment be identical. 
The internal structure of the environment degree of freedom does not enter any of the requirements we derive for quantum Darwinism to emerge, therefore it is unnecessary to assume anything about it.
 
Leaving any free Hamiltonian acting on the system $\sn{S}$ or any environment degree of freedom $\sn{E}_j$ unspecified for now, the most general Hamiltonian   includes arbitrary two-body interactions between the system and environment,
\begin{align}
    \HH = \Sop{\HH_\sn{S}} + \Eop{\HH_\sn{E}} + \sum_{j=1}^{|\sn{E}|}\sum_{k=1}^{|\mathcal{L}(\sn{E}_j)|} \opr{S}_{jk}\otimes\opr{E}_{jk}
    ,
    \label{eqn:GenericHamiltonian}
\end{align}
where $j$ indexes the environment degrees of freedom and $k$ indexes some basis of operators acting on the $j$th environment subsystem with size $|\mathcal{L}(\sn{E}_j)|$.
The system and environment operators $\opr{S}_{jk}$ and $\opr{E}_{jk}$ are taken to be traceless.
 
Following the results of Ref.~\cite{Duruisseau23}, we will assume that each $\opr{S}_{jk}$ contains within a random prefactor which is drawn from some random distribution with continuous support. 
In the qubit case \cite{Duruisseau23}, an exact solution of the dynamics was possible which showed that distributions with discrete support do not \emph{asymptotically} support quantum Darwinism, since any information transfer is bidirectional and periodic.
While it is not possible to produce an exact solution of the dynamics of the generic model we consider in this work, we expect that in most circumstances the same requirement applies.
We will revisit this statement when we discuss the possibility of relaxing the constraints we construct in the remainder of this section.
 
Now, we may translate the requirements of quantum Darwinism into statements about Hamiltonians of the form shown in Eq.~\eqref{eqn:GenericHamiltonian} if they are to exhibit quantum Darwinism. 

\subsection{Pointer observable}
As stated in the introduction, the most basic requirement for quantum Darwinism is that there must exist some pointer observable $\opr{\mathcal{A}} = \Sop{\opr{A}}$ which commutes with the overall Hamiltonian \cite{Duruisseau23,Zurek81,Zurek82,Ollivier05},
\begin{align}
    \comm{\opr{\mathcal{A}}}{\HH} = 0
    .
\end{align}
The eigenstates of this operator (the ``pointer states'') are those states that remain stationary under the joint system-environment interaction and into which the system ultimately decoheres due to the interaction with the environment.
 
For the Hamiltonian of Eq.~\eqref{eqn:GenericHamiltonian} to support a pointer basis, there must exist some operator $\opr{A}$ such that
\begin{equation}
    \comm{\opr{A}}{\HH_\sn{S}} = 0 \quad\text{and}\quad  \comm{\opr{A}}{\opr{S}_{jk}}= 0 \qquad \forall j,k\,
    .
    \label{eqn:PointerBasisReqs}
\end{equation}
In general, fixing the system Hamiltonian $\HH_\sn{S}$ does not uniquely specify a pointer basis.
\footnote{In the qubit model the requirements of Eq.~\eqref{eqn:PointerBasisReqs} are sufficient to fix both $\opr{A}$ and $\opr{S}_{jk}$ and derive Eq.~\eqref{eqn:QBModelH} once the system Hamiltonian $\HH_\sn{S}$ is given.
This is because in the space of qubit operators, only the identity and constant multiples of $\HH_\sn{S}$ commute with $\HH_\sn{S}$.}
 
If the system is larger than a qubit, then the existence of a pointer basis does \emph{not} necessarily require that the system operators entering the interaction commute with each other nor does it require that they commute with the free Hamiltonian.
This is possible if $\opr{A}$ is degenerate, in which case it is the projections onto the degenerate decoherence-free subspaces of $\opr{A}$ which are preserved by the dynamics \cite{Zurek03,Zurek13}. 
Degenerate pointer observables indicate that only partial information about the system state is available to be encoded into the environment and that the decohered system density matrix is not necessarily purely diagonal but may have some persistent coherence. 
 
As we will see in the next section, however, while it is possible for a Hamiltonian with non-commuting interactions to support a pointer observable we find the information about that observable is not encoded in the environment in a redundant and accessible way and therefore there is no quantum Darwinism in such a case.

\subsection{Singly-branching form}
\label{ssec:SinglyBranchingForm}
Once the notion of objectivity is defined in terms of indirect measurements through environment fragments as is done in quantum Darwinism and related ideas such as strong quantum Darwinism and the spectrum broadcast structures approach, it can be shown that the \emph{only} system-environment states which support objectivity are states of singly-branching form \cite{Touil22}, i.e., states of the form,
\begin{align}
    \ket{\psi(t)} = \sum_{n} c_n \ket{n}\bigotimes_{j=1}^{|\sn{E}|}\ket{\phi_{nj}(t)}
\end{align}
where the index $n$ runs over the different pointer states $\ket{n}$ on which the system state has support and where $\ket*{\phi_{nj}(t)}$ denotes the state of the $j$th environment degree of freedom conditioned on the system state being $\ket{n}$ (which can be generalized to density operators if necessary).
The distinguishability of the conditional states $\ket*{\phi_{nj}(t)}$ and $\ket*{\phi_{n'j}(t)}$ are what allow the observer to learn about the coefficients $c_n$ from measurements on environment fragments, and the lack of additional correlations between environment degrees of freedom allows multiple observers to make independent measurements. 
In the specific case where the conditional states are orthogonal this is an example of a spectrum broadcast structure \cite{Horodecki15,Korbicz21}.

For our purposes, where we are interested in the asymptotic limit of very large environments, it would be overly-restrictive to require that the conditional states of the environment degrees of freedom be exactly orthogonal.
So long as the conditional states are distinguishable, then the conditional states of sufficiently large (but still tiny compared to the full environment) environment fragments are almost orthogonal \cite{Zurek22}.
 
The most obvious restriction on $\HH$ that arises from the need to preserve the singly-branching form is that there can be no intra-environment interactions in the free Hamiltonian describing the environment, i.e. that we may decompose
\begin{align}
    \HH_\sn{E} = \sum_{j=1}^{|\sn{E}|} \HH_{\sn{E}_j}
    ,
    \label{eqn:LocalEnvH}
\end{align}
into a collection of free Hamiltonians each acting on a single environment degree of freedom.
Any interactions in $\HH_\sn{E}$, e.g. $\opr{E}_i\otimes\opr{E}_j$, would take a product state $\ket{\psi_i}\otimes\ket{\psi_j}$ to some entangled state under time evolution, which is incompatible with the singly-branching form.
 
In the qubit case, this requirement that there be no intra-environmental interactions was the only condition necessary to exclude mixing in the environment since the form of the system-environment interaction was fixed by the pointer basis requirement \cite{Duruisseau23}.
In the general case there are additional conditions which must be satisfied by $\HH$. We summarize these requirements and their origin here, the details of their derivation are presented in Appendix~\ref{app:DeriveHamiltoniansPointerBasis}.

\subsubsection{Induced intra-environmental mixing}
Suppose we have a Hamiltonian of the form given in Eq.~\eqref{eqn:GenericHamiltonian} which supports a pointer basis, i.e. there exists some pointer observable $\opr{A}$ which commutes with $\HH_\sn{S}$ and each $\opr{S}_{jk}$.
Clearly, the time evolution of the joint system-environment state described by the propagator
\begin{align}
    \opr{U}(t) = e^{-it \HH}
    ,
\end{align}
preserves the pointer states of the system.
However, it is possible that the components of $\HH_\sn{S}$, $\VV_\sn{SE}$, and/or $\HH_\sn{E}$ do not commute with one another.
Therefore we can not factor the propagator trivially into three pieces as we would like to easily understand the evolution of singly branching states, instead we expand the exponential using the Zassenhaus formula \cite{Casas12} 
\begin{align}
    e^{A+B} = 
        e^A 
        e^B
        e^{-\frac{1}{2}\comm{A}{B}}
        e^{\frac{1}{3}\comm{B}{\comm{A}{B}} + \frac{1}{6}\comm{A}{\comm{A}{B}}}
        \dots
\end{align}
into an infinite product of propagators.
 
Consider the case where different interactions in $\VV_\sn{SE}$ do not commute, meaning that there exist some pairs of system operators such that $\comm{\opr{S}_{j'k'}}{\opr{S}_{jk}} \ne 0$.
Then, at second order in the expansion of the full propagator there are terms of the form
\begin{align}
    \comm{\opr{S}_{j'k'}}{\opr{S}_{jk}} \otimes \opr{E}_{j'k'} \otimes \opr{E}_{jk}
    ,
    \label{eqn:SSComm}
\end{align}
and terms with similar structures at higher orders. 
If $j\ne j'$, then $\opr{E}_{jk}$ and $\opr{E}_{j'k'}$ are operators acting on different environment degrees of freedom and so this term represents an effective interaction between the environment degrees of freedom $j$ and $j'$ mediated by the system. 
This can -- and as we will show in an example model does -- lead to mixing dynamics in the environment incompatible with quantum Darwinism.
Therefore, we must require that the system operators entering the interaction Hamiltonian commute with one another.
Not to ensure the existence of a pointer basis, but to ensure that the information about the system in that pointer basis is redundantly and accessibly encoded in the environment.
 
In fact we must impose a somewhat stronger version of this requirement once we consider the possibility that the free system Hamiltonian $\HH_\sn{S}$ could also fail to commute with one or more of the interaction operators while still preserving a pointer basis. 
Again, by expanding the propagator (this time to third order) we find terms of the form
\begin{align}
    \comm{\opr{S}_{j'k'}}{\comm{\HH_\sn{S}}{\opr{S}_{jk}}}\otimes\opr{E}_{j'k'} \otimes \opr{E}_{jk}
    ,
    \label{eqn:SHSComm}
\end{align}
representing another effective intra-environment interaction mediated by the system if $j \ne j'$.
Thus, we also must require that these commutators vanish if $\HH$ is to support quantum Darwinism.
 
The timescales at which these higher-order processes induce mixing between the environment sites may be long, and indeed it may be possible to observe some level of objectivity at short or intermediate times before the mixing dynamics has destroyed the approximate singly-branching form of the system-environment state vector supported by lower-order terms \cite{Duruisseau23,Riedel12}.
In the long-time asymptotic regime most relevant to macroscopic objectivity, however, eventually all orders of the expansion of the propagator become relevant and so we must require that this ``non-mixing'' property hold at every order.
 
To simplify the notation, we introduce a set of commutator superoperators $\mathcal{C}_\mu$ defined such that
\begin{align}
    \mathcal{C}_0 \opr{O} \equiv \comm{\HH_\sn{S}}{\opr{O}}
    ,
    \qquad
    \mathcal{C}_{jk} \opr{O} \equiv \comm{\opr{S}_{jk}}{\opr{O}}
    .
    \label{eqn:CommutatorSops}
\end{align}
If we additionally define $\tilde{E}_\mu$ such that 
\begin{align}
    \tilde{E}_0 = \bigotimes_{n=1}^{|\sn{E}|} \mathbb{I}_n
    ,
    \qquad
    \tilde{E}_{jk} = \opr{E}_{jk}\!\!\bigotimes_{n=1; n\ne j}^{|\sn{E}|}\!\!\mathbb{I}_n
    ,
    \label{eqn:ETilde}
\end{align}
then all the nested commutators in the expansion can be written as
\begin{align}
    \pqty{\prod_{n=1}^{|\mu|} \mathcal{C}_{\mu(n)}} \opr{S}_{jk}
    \otimes
    \pqty{\prod_{n=1}^{|\mu|} \tilde{E}_{\mu(n)}} \tilde{E}_{jk}
    ,
    \label{eqn:NestedZassenhaus}
\end{align}
where $\mu$ is a sequence of length $|\mu|$ whose entries $\mu(n) \in \{0\}\cup(j,k)$ with $j\in\{1,\dots,|\sn{E}|\}$, $k\in\{1,\dots,|\mathcal{L}(\sn{E}_j)|\}$ indicate which commutator superoperator is to be applied.
We arbitrarily define the products as growing to the left ($\dots\mathcal{C}_{\mu(2)}\mathcal{C}_{\mu(1)}$) so that the sequence $\mu$ more obviously matches the recursive structure of the nested commutators.
For example, the zeroth-order contribution corresponds to an empty sequence, Eq.~\eqref{eqn:SSComm} corresponds to the length-$1$ sequence with $\mu(1) = (j',k')$, and Eq.~\eqref{eqn:SHSComm} to the length-$2$ sequence with $\mu(1) = 0$, $\mu(2) = (j',k')$.
 
Clearly, if there is any element in the sequence $\mu$ such that $\mu(m) = (j',k') \ne 0$ with $j'\ne j$ then the product of environment operators in Eq.~\eqref{eqn:NestedZassenhaus} represents an intra-environmental interaction between $\sn{E}_j$ and $\sn{E}_{j'}$.
The same is true if the sequence contains any two elements $(j',k')$ and $(j'',k'')$ with $j'\ne j''$.
Therefore, to exclude any potential mixing dynamics, the associated nested commutator of system operators must vanish for any sequence $\mu$ with either of these properties,
\begin{align}
    \pqty{\prod_{n=1}^{|\mu|} \mathcal{C}_{\mu(n)}} \opr{S}_{jk} = 0
    \qquad
    \forall j,k,\mu
    .
    \label{eqn:SopStringZero}
\end{align}
 
Note that if all non-zero elements of the sequence of indices have the same $j'=j$, then the operator from Eq.~\eqref{eqn:NestedZassenhaus} does not represent an effective interaction and so would support the singly-branching form.
This is not a particularly interesting scenario when the system is finite dimensional, however, since then it is only possible for the system to interact with finitely-many environment degrees of freedom in this way (limited by $|\mathcal{L}(\sn{S})|$).
When taking the limit of a large environment, rather than consider the few degrees of freedom which interact with the system in this way as part of the large environment it is more reasonable to interpret them as either a second small finite environment or as part of the system itself.
Thus, we do not pursue models with this behavior.
 
What remains to consider is the environment Hamiltonian, which is much simpler.
As discussed before, we have already excluded the possibility of direct interactions between distinct degrees of freedom in $\HH_\sn{E}$ by requiring that the environment Hamiltonian be of the form presented in Eq.~\eqref{eqn:LocalEnvH}.
As for the interaction $\VV_\sn{SE}$, by construction we have that every $\opr{E}_{jk}$ commutes with every other $\opr{E}_{j'k'}$ (if $j\ne j'$ they correspond to different degrees of freedom, or if $j=j'$ but $k\ne k'$ to different operators in the linearly independent basis for $\mathcal{L}(\sn{E}_j)$).
It is possible that $\HH_\sn{E_j}$ fails to commute with any $\opr{E}_{jk}$, however this can not affect either the existence of a pointer basis or the preservation of the singly-branching form as such commutators cannot induce interactions. 
It is possible that the timescale for the emergence of classical objectivity is affected or that the minimum size of the environment fragment $|\sn{F}|$ from which the full accessible information about the system can be extracted grows, but in the limit of sufficiently long times and sufficiently large environments objectivity with emerge nonetheless.

\subsubsection{Relaxing these constraints}
Before concluding this discussion, we should note that the criteria we have placed on the Hamiltonian in order to exclude mixing terms and therefore preserve the singly branching form required for quantum Darwinism are in some sense overly restrictive.
 
From an intuitive standpoint it is necessary to exclude intra-environmental interactions as they lead to the information about the system which is encoded into each degree of freedom spreading throughout the environment, becoming non-local and inaccessible to measurements on small fragments of the environment.
However, consider a scenario including an interaction Hamiltonian which only includes interactions between disjoint pairs of environment degrees of freedom, e.g.,
\begin{align}
    \VV_{\sn{E}\sn{E}} = \sum_{j=1}^{\lfloor|\sn{E}|/2\rfloor} \opr{B}_{2j}\otimes\opr{B}_{2j-1}
    .
\end{align}
In this case, if we redefine the environment $\sn{E}\to\widetilde{\sn{E}}$ such that the $\ell$th degree of freedom in the new environment corresponds to the composite subsystem $\sn{E}_{2\ell}\otimes\sn{E}_{2\ell-1}$ in the original environment, then we can see the emergence of quantum Darwinism in this model.
For example, in Ref.~\cite{Chisholm22} a system-environment model was studied where the system interacts with interacting pairs of environment qubits.
If the environment fragments are required to consist of interacting pairs, the mutual information exhibits the characteristic plateau whereas if the fragments are allowed to be arbitrary, instead the mutual information plot shows no redundancy.
 
Even without this redefinition it may be possible to see the emergence of objectivity in the presence of intra-environmental interactions so long as they are sufficiently sparse. For example, imagine introducing a coupling between exactly two environment degrees of freedom in a very large environment. For this coupling to make the system information inaccessible to any given fragment, that fragment (i) must include one and only one of the two ``bad'' degrees of freedom and (ii) must be very small, else the other degrees of freedom in the fragment themselves contain enough redundant information to recover the system information.
The joint system-environment state is clearly very close to the singly-branching form, hence the quantum discord is very close to zero and so this model demonstrates emergent quantum Darwinism \cite{Duruisseau23}.
Exactly how numerous and how strong intra-environmental interactions can be while still supporting quantum Darwinism with nearly-singly-branching states is an interesting question, however it is beyond the scope of the present work \footnote{For instance, in Ref.~\cite{Touil22} it is shown how the strict adherence to single-branching states can be relaxed to ``$\epsilon$-$\delta$-statements.''}. 
 
Since we were not required to make any assumptions about the internal structure about each environment degree of freedom in our model, we choose to interpret scenarios where there is mixing between disjoint subsets of the environment in terms of this replacement. 
This is not an issue from the point of view of an observer measuring the environment, we are merely telling the observer that when deciding on a fragment of the environment to measure they should choose to include either all or none of any set of interacting degrees of freedom.
Of course, if large subsets of the environment are mutually interacting then after this redefinition there are only a few effective degrees of freedom and the redundancy of the system information is low as should be expected. 

Another practically-motivated approach to loosening the constraints on mixing between environment degrees of freedom would be to relax the requirement that quantum Darwinism persist as $t\to\infty$. 
For example, as will be visible in some of the following examples and as has been shown in other works \cite{Touil22,Riedel12}, in a model where intra-environmental mixing is present but very weak we expect to see the system-environment state flow to an approximately singly-branching state with its associated redundancy and objectivity after a short initial transient. Eventually, there will be a very slow relaxation of the joint state to a typical entangled state with no redundancy. 
For some purposes this sort of long-lived but ultimately temporary quantum Darwinism may be sufficient, e.g. if one were to consider a scenario where there is some upper limit on the time available for observers to make measurements.
However, it is clear that such models must in a certain sense be close to models which asymptotically exhibit quantum Darwinism, and hence we focus on the asymptotic case in this work.
 
Before we consider any explicit examples, we will revisit our requirement that the system operators $\opr{S}_{jk}$ entering the interaction be proportional to random prefactors which are drawn from a distribution with continuous support.
In the simple qubit model of Ref.~\cite{Duruisseau23}, this requirement can be derived from a straightforward analytical solution for the exact system-environment dynamics.
The mutual information between environment fragments and the system is related to the average overlap of the conditional environment states corresponding to the pointer states, $\ev{|\Gamma(t)|^2}$, called the decoherence factor.
If $\ev{|\Gamma(t\to\infty)|^2} \to \epsilon < 1$, then at sufficiently long times the system-environment dynamics produce a stable plateau in the mutual information curve indicative of quantum Darwinism.  
For the qubit model, the functional form of the average decoherence factor is
\begin{align}
    \ev{|\Gamma(t)|^2} = a + b\Re\{\mathcal{F}(k t)\}
    \label{eqn:DecoherenceFactorFT}
\end{align}
where $a,b,k\in\mathbb{R}$ are constants with $a+b=1$ and $a<1$, and $\mathcal{F}$ is the Fourier transform of the probability density function of the coefficients. 
Asymptotically, the term including $\mathcal{F}(kt)$ vanishes if and only if the distribution has continuous support.
If it does not, then $\ev{|\Gamma(t)|^2}$ is a periodic or quasi-periodic function of time, and so for any cutoff time $\tau$ there is eventually always some $t>\tau$ such that the decoherence factor becomes arbitrarily close to one, which coincides with the classical plateau in the mutual information plot vanishing.
This may only occur at extremely long times, especially if the environment is very large and the average decoherence factor is quasi-periodic, but it eventually will -- the information transfer is \emph{not} irreversible and the emergent notion of objectivity is not stable.
This may be a sufficient substitute in some cases, however in this work we are interested in the asymptotic emergence of quantum Darwinism which is persistent. 
 
For an arbitrary finite-dimensional system and environment model, the behavior of the decoherence factor is the same. 
The single evaluation of the Fourier transform in Eq.~\eqref{eqn:DecoherenceFactorFT} generally expands to a large and complicated function involving the evaluation of the Fourier transforms of all coefficient distributions at arguments spaced related to the natural frequency scales in the model, which may or may not be commensurate.
In the asymptotic limit $t\to\infty$, all the sensitivity on the coefficient distributions drops out and the decoherence factor reduces to a constant if all the transforms vanish at sufficiently high frequencies, i.e., if the distributions have continuous support.

\subsubsection{Correlations in the initial state}
Our discussion on the necessity of system-environment states of singly-branching form for objectivity has thus far been focused on the preservation of such states dynamically and thus on the requirements imposed on Hamiltonians which support quantum Darwinism. 
If the initial system-environment state is a product state or more generally a state already of the required singly-branching form, then the evolution under such a Hamiltonian will lead to the emergence of quantum Darwinism (or its preservation, in the case the initial state already exhibits objectivity).

Suppose instead that the initial state is not of singly-branching form, meaning that it includes correlations between environment degrees of freedom.
Since a Hamiltonian which supports quantum Darwinism must have no intra-environment interactions, even effectively, these initial correlations will never be destroyed through time evolution. 
Ensuring that the Hamiltonian can never generate these quantum correlations in the environment also ensures that such correlations are ``invisible'' to the Hamiltonian.
To observers, however, the correlations are extremely important and their presence prevents the emergence of redundancy and hence objectivity. 

To conclude this discussion, we have identified one final constraint necessary for a model to exhibit quantum Darwinism, though not a constraint on the Hamiltonian itself: the initial system-environment state must not posses quantum correlations in the environment.
This should not be a particularly stringent constraint in practice, as any such correlations would have to have been created with a different Hamiltonian which does not support quantum Darwinism. 
We also note that, just as was the case with the presence of intra-environmental interactions in the Hamiltonian discussed above, depending on the details of the initial correlations it may be possible to coarse-grain the environment into uncorrelated effective degrees of freedom.

\section{Time-independent examples}
\label{sec:Qutrit}
%

Having discussed the requirements which must be placed on a generic Hamiltonian to support quantum Darwinism, we now present a paradigmatic example which shows how Hamiltonians more general than those required in qubit models may be used.
 
As a first example, we consider is a small modification to the qubit model \cite{Duruisseau23}, where now a single three-level qutrit system interacts with an environment of qubits with two-body interactions between the system and environment and no intra-environment interactions.
The corresponding Hamiltonian is
\begin{align}
    \HH &= 
        \Sop{\sum_{\gamma}B^\gamma\opr{O}^\gamma} + \Eop{\sum_{j,\beta}B^\beta_j\opr{\sigma}^\beta_j} 
        \nonumber\\
        &\quad+ 
        \sum_{\gamma}\sum_{j,\beta}J^{\beta\gamma}_j\opr{O}^\gamma\otimes\opr{\sigma}^\beta_j
\end{align}
where $B^\beta_j$ and $J^{\beta\gamma}_j$ are randomly selected coefficients, $j$ indexes environment qubits, $\sigma^\beta$ the Pauli operators, and $\opr{O}^\gamma$ represents a qutrit operator corresponding to one of the Gell-Mann matrices \cite{Griffiths}.
 
To explore the space of behaviors, we study four example models numerically. 
The first three of the example models we consider all support quantum Darwinism, but they move increasingly away from the separable Hamiltonians required for a simpler qubit model to support quantum Darwinism.
The final example fails to support a pointer basis and so fails to exhibit any emergence of classicality.

For simplicity, in each of these models we set all local Hamiltonians to zero (i.e., $B^\gamma = B^\beta_j = 0~~\forall j$) as we are more interested in the structure of the interaction.
As discussed in Appendix~\ref{app:DeriveHamiltoniansPointerBasis}, the local Hamiltonians on each environment degree of freedom are essentially irrelevant for the asymptotic emergence of quantum Darwinism -- they can only affect the rate of emergence. 
A local system Hamiltonian, conversely, is quite impactful due to the constraints given in Eq.~\eqref{eqn:SopStringZero} placing restrictions on what interaction Hamiltonians can be allowed while still yielding quantum Darwinism. 
An example of the effects of ``incompatibility'' between the system Hamiltonian and interaction in a qubit model, is given in Ref.~\cite{Duruisseau23}. 
As will be demonstrated, the behavior of that qubit model is qualitatively similar to our fourth model, demonstrating the inherent similarity in the underlying mechanism by which quantum Darwinism is obstructed. 

\begin{figure*}
    \centering
    \includegraphics[width=\textwidth]{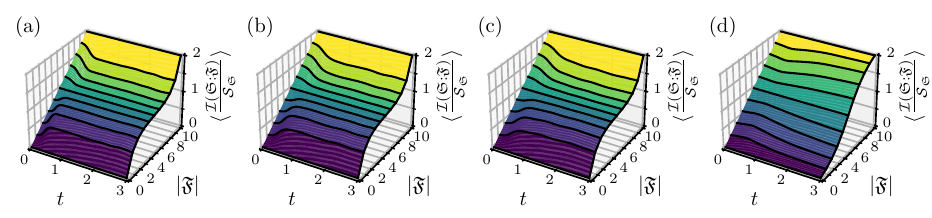}
    \caption{Evolution of the normalized mutual information between the system qutrit and environment fragments of a given size $\sn{F}$ as a function of time, for (a) model A, (b) model B, (c) model C, (d) model D. Each plot shows the mutual information averaged over $100$ simulations with randomized choices of coefficients for each Hamiltonian, all starting from the initial state of Eq.~\eqref{eqn:QtInitialState}.}
    \label{fig:QutritQDExamples}
\end{figure*}

\subsubsection{Model A: Globally separable}
The first model is described by a separable system-environment interaction, where the system interacts with each environment degree of freedom equivalently.
The Hamiltonian we choose for this case is
\begin{align}
    \HH_A = \pqty{\opr{Z}_2 + \opr{X}_{01}} \otimes \sum_{i=1}^N J_i \sigz_i 
    ,
    \label{eqn:QutritHGloballySep}
\end{align}
where the coefficients $J_i\sim\mathcal{N}(0,\sigma_J^2)$ are taken from a normal distribution with zero mean.
In the number basis, the qutrit operators are
\begin{subequations}\begin{align}
    \opr{Z}_2 &= \big(
        \dyad{0} + \dyad{1} - 2\dyad{2}
    \big)/\sqrt{3}
    \\
    \opr{X}_{01} &= \big(
        \dyad{0}{1} + \dyad{1}{0}
    \big)
    .
\end{align}\end{subequations}
The qutrit operator $(Z_2 + X_{01})$ is non-degenerate, therefore its eigenstates
\begin{subequations}
\begin{align}
    \ket{p_0} &= \pqty{\ket{0} + \ket{1}}/\sqrt{2},
    \\
    \ket{p_1} &= \pqty{\ket{0} - \ket{1}}/\sqrt{2},
    \\
    \ket{p_2} &= \ket{2}
    .
\end{align}
    \label{eqn:QtPointerBasis}
\end{subequations}
are the pointer states defining the basis in which the system dephases.
 
Figure~\ref{fig:QutritQDExamples}(a) shows the evolution of the normalized mutual information between environment fragments of different sizes and the system as a function of time, averaged over $100$ simulations with random choices of coefficients with $\sigma_J = 1$ and $|\sn{E}| = 10$ qubits. 
In each case, the initial system-environment state is a product state, 
\begin{align}
    \ket{\psi(0)} = \frac{\ket{0}+\ket{1}+\ket{2}}{\sqrt{3}}\bigotimes_{i=1}^N \frac{\ket{0_i}+\ket{1_i}}{\sqrt{2}}
    \label{eqn:QtInitialState}
\end{align}
We observe the rapid development of the classical plateau emblematic of quantum Darwinism, as we expect in this system.

It should be noted that the averaging over $100$ simulations performed to construct Fig.~\ref{fig:QutritQDExamples}(a) is a computational technique which we employ due to the small environment sizes directly accessible with our simulations, and is not related to quantum Darwinism. 
If we could simulate large enough environments, we would observe decoherence of the system and emergence of the mutual information plateau in each individual trial.
For simple models such as the one just considered or the simple models of Ref.~\cite{Duruisseau23} where the dynamics can be solved analytically, it can be shown that this averaging procedure correctly captures the large-$N$ behavior (up to minor details related to rescaling of interaction strengths and the width of the plateau as $N\to\infty$) while keeping the actual simulations tractable. 
For more complicated models which cannot be analytically solved we nonetheless expect similarly accurate results, again up to some minor details related to rescaling of quantities with $N$. 

\subsubsection{Model B: Non-separable, commuting environment}
The second example replaces the fixed system operator of the previous model with a collection of random qutrit operators which all support the same pointer basis of Eq.~\eqref{eqn:QtPointerBasis}. 
Explicitly, we consider the Hamiltonian
\begin{align}
    \HH_B &= 
        \sum_{i=1}^N \pqty{J_i\opr{Z}_2 + K_i\opr{X}_{01}}\otimes\sigz_i 
        \nonumber\\
        &=
        \opr{Z}_2\otimes\sum_{i=1}^N J_i \sigz_i + \opr{X}_{01}\otimes\sum_{i=1}^N K_i \sigz_i
    ,
\end{align}
where the coefficients are drawn from independent normal distributions with zero mean, $J_i \sim \mathcal{N}(0,\sigma_J^2)$ and $K_i \sim \mathcal{N}(0,\sigma_K^2)$. 
The interaction between the system and each environment qubit is still separable, but now the overall system-environment interaction is non-separable.
 
As this Hamiltonian supports exactly the same pointer basis as the previous example, we expect the dynamics to be very similar in the two situations.
This is borne out in Fig.~\ref{fig:QutritQDExamples}(b) which shows that this model exhibits essentially the same behavior as the previous example.
The timescale at which the classical plateau emerges is slightly slower than before, but it still rapidly and convincingly appears.

\subsubsection{Model C: Non-separable, non-commuting environment}
Our third example modifies the previous example case by choosing to have the system operators $\opr{Z}_2$ and $\opr{X}_{01}$ couple to non-commuting operators acting on each environment qubit, according to the Hamiltonian
\begin{align}
    \HH_C &= 
        \sum_{i=1}^N \pqty{J_i\opr{Z}_2 \otimes \sigz_i + K_i\opr{X}_{01} \otimes \sigx_i}
        \nonumber\\
        &=
        \opr{Z}_2\otimes\sum_{i=1}^N J_i \sigz_i + \opr{X}_{01}\otimes\sum_{i=1}^N K_i \sigx_i
    ,
\end{align}
where the coefficients are distributed as in the previous example. 
Note that in this example, not only is the overall system-environment interaction not separable but the interaction between the system and each environment qubit is not separable.
 
Nonetheless, this fact has no bearing on the ability of this model to support quantum Darwinism.
The same pointer states from Eq.~\eqref{eqn:QtPointerBasis} are stationary under the Hamiltonian, and there is no possibility of induced intra-environmental mixing thus we should expect to see objectivity emerge.
As we can see in Fig.~\ref{fig:QutritQDExamples}(c), our expectation is borne out and this model exhibits quantum Darwinism just as the previous two models did. 
The rate at which objectivity emerges is again slightly slower, but this is unimportant.
 
Before continuing to our last example, we summarize an important message from the three examples considered thus far: the (non-)separability of the system-environment interaction is itself not relevant as to whether or not any given Hamiltonian supports quantum Darwinism.
Only the necessity of the existence of pointer states and system-environment states of singly-branching form.

\subsubsection{Model D: Non-commuting system operators}
The final example   is
\begin{align}
    \HH_D = \sum_{i=1}^N J_i\opr{X}_{02} \otimes \sigz_i + \sum_{i=1}^N K_i\opr{X}_{01} \otimes \sigx_i
    ,
    \label{eqn:QutritHBad}
\end{align}
again with normally distributed coefficients, which fails to exhibit quantum Darwinism as shown in Fig.~\ref{fig:QutritQDExamples}(d) since there exists no pointer basis which is preserved by both $\opr{X}_{01}$ and $\opr{X}_{02}$.
This is exactly what we expect, since the failure of the system operators entering the interaction violate our criteria.
 
\begin{figure}[t]
    \centering
    \includegraphics[width=0.9\columnwidth]{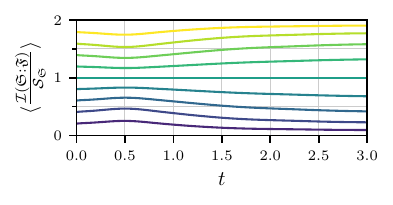}
    \caption{Alternate view of Fig.~\ref{fig:QutritQDExamples}(d) showing the average mutual information fragments of size $1\le|\sn{F}|\le9$ evolving under the Hamiltonian of model D. Beyond $t\approx0.5$ the mixing terms induced by the system-environment interaction cause the information encoded in the environment to spread, precluding the emergence of objectivity in this case.}
    \label{fig:QutritQDResD}
\end{figure}

Interestingly, at short times ($t<0.5$) we see the information content of the different fragment sizes all tending towards $1$, as we would expect for a case which exhibits quantum Darwinism. 
Figure~\ref{fig:QutritQDResD} shows a different view of how the average normalized mutual information varies over time for different sizes of the environment fraction under this model which makes this more visible.
It is only at longer times ($t>0.5$) that this is reversed and the mutual information of fragments consisting of less than half the environment shrink toward zero and of those consisting of more than half the environment toward two, as would be expected for a Hamiltonian which includes scrambling dynamics in the environment.
The finite size of the simulated environment limits the magnitude of these two effects, but they are clearly visible in the data nonetheless. 
 
This result matches with our understanding of where the scrambling in this model comes from. 
At very short times, only the lowest order contribution to the Zassenhaus expansion \eqref{eqn:NestedZassenhaus} of the propagator is relevant, and this contribution does not include any mixing. 
It is only at later times that the higher order contributions with their induced intra-environmental interactions become dominant, leading to the observed asymptotic behavior.
It is interesting to contrast the behavior of this model with the ``CODI'' qubit model presented in Ref.~\cite{Duruisseau23} containing a non-commuting local Hamiltonian acting on the system. 
The evolution of the mutual information for the two models are extremely similar, demonstrating that there is little qualitative difference between the present model with no pointer basis due to non-commuting interactions and a model with no pointer basis due to a non-commuting system Hamiltonian.

\section{Time-dependent model}
\label{sec:TimeDep}
%

In this section, we will consider a further generalization of the model of Sec.~\ref{sec:Model} to encompass arbitrary time-dependent free and interaction Hamiltonians.
The generalization of the Hamiltonian given in Eq.~\eqref{eqn:GenericHamiltonian} is straightforward, the only difference is that now we allow every operator entering the Hamiltonian to be arbitrarily time-dependent,
\begin{align}
    \HH(t) 
    &= \Sop{\HH_\sn{S}(t)} + \Eop{\HH_\sn{E}(t)} 
    \nonumber\\
    &\qquad+ \sum_{j=1}^{|\sn{E}|}\sum_{k=1}^{|\mathcal{L}(\sn{E}_j)|} \opr{S}_{jk}(t)\otimes\opr{E}_{jk}(t)
    ,
    \label{eqn:GenericTdHamiltonian}
\end{align}
where as before $j$ indexes the environment degrees of freedom $\sn{E}_j$ and $k$ indexes a basis of operators acting on that subsystem and we assume the operators entering the interaction to be traceless and proportional to some random coefficient drawn from a distribution with continuous support.
 
We will assume that there is no cutoff time $T$ which partitions the environment degrees of freedom such that the system only interacts with some $\sn{E}_j$ before $T$ and some disjoint set after.
If there were, we could consider the evolution from $t=0$ to $T$ to be essentially part of a state preparation protocol and ignore the $\sn{E}_j$ which only interact in that time when building environment fragments to measure. 
Then, after this initial preparation process, quantum Darwinism would emerge as expected but with the information about the perturbed system state being encoded into the environment.
Note that while the rate at which information is encoded into the environment for times $t>T$ could vary depending on the details of the effective preparation protocol, this would be simply due to asymmetries in the Hamiltonian causing different subspaces of the pointer observable decohering at different rates. 
Such effects may be present in all models of quantum Darwinism, with or without time dependence.

We additionally assume that the system continues to interact with the environment even in the $t\to\infty$ limit, e.g. we do not consider interactions with a prefactor $\propto\exp(-t/\tau)$ for some $\tau$.
We leave studying the possibility of the asymptotic emergence of quantum Darwinism or some similar approximate notion in asymptotically decoupled models as an avenue for future study. 
 
From the point of view of an observer searching for objectivity, there are no substantive changes required to consider a time-dependent system.
Objectivity still requires the existence of some pointer observable $\opr{\mathcal{A}} = \Sop{\opr{A}}$ which defines the pointer states, the information about which must be encoded redundantly and accessibly in the environment degrees of freedom.
Therefore, just as before, there must exist some $\opr{A}$ such that
\begin{align}
    \comm{\opr{A}}{\HH_\sn{S}(t)} 
    =
    \comm{\opr{A}}{\opr{S}_{jk}(t)} 
    = 0
    .
\end{align}
Note that it is critical that the pointer observable is a time-\emph{independent} operator -- it defines the set of pointer states that are preserved by the interaction and which identify the information about the system that becomes objective, and that information must be stable to be redundantly encoded in the environment. 
While scenarios may exist where the system-environment interaction supports a pointer observable with some predictable time dependence such that observers could still reconstruct system information from environment fragments, that would certainly not fall under the precise notions of objectivity and quantum Darwinism we take in this work.
Hence, we require that the pointer observable remain perfectly stationary in time.

Next, again just as in the time-independent case, we must require that $\HH(t)$ preserve the singly-branching form.
At minimum, this requires that the environment Hamiltonian can be written purely in terms of local Hamiltonians acting on each subsystem independently,
\begin{align}
    \HH_\sn{E}(t) = \sum_{j=1}^{|\sn{E}|} \HH_{\sn{E}_j}(t)
    ,
\end{align}
such that there is no explicit intra-environment scrambling dynamics present.
 
There are further constraints on the system operators entering the interaction which must be satisfied to avoid the introduction of intra-environment mixing mediated by the system.
We summarize these constraints here, for details see Appendix~\ref{app:DeriveTdHamiltonians}.
The approach is broadly similar to the time-independent case, beginning with taking the propagator for the time evolution generated by the full Hamiltonian $\HH(t)$ and decomposing it into an infinite product of increasingly higher-order propagators.
In doing so, we find terms with time-dependent generalizations of the nested commutator structures seen in the time-independent case such as (cf. Eq.~\eqref{eqn:SSComm}, similarly Eq.~\eqref{eqn:SHSComm})
\begin{align}
    \int_{0}^{t}\dd{t_1}\int_{0}^{t_1}\dd{t_0}\comm{\opr{S}_{j'k'}(t_1)}{\opr{S}_{jk}(t_0)}\otimes \opr{E}_{j'k'}(t_1) \opr{E}_{jk}(t_0)
    .
    \label{eqn:FirstTdMixingConstraint}
\end{align}
Such terms lead to mixing if the commutator is not zero for all times $t_1,t_2$, if the environment operators act on different degrees of freedom.
 
These and similar terms compel us to pose a straightforward generalization of the restriction placed on commutators of the system operators in the time-independent case. 
After defining the now time-dependent commutator superoperators $\mathcal{C}_\mu(t)$ and environment operators $\tilde{E}_{\mu}(t)$ (cf. Eq.~\eqref{eqn:CommutatorSops} and Eq.~\eqref{eqn:ETilde}), the various commutators which are integrated over in the expansion can be written as,
\begin{align}
    \pqty{\prod_{n=1}^{|\mu|} \mathcal{C}_{\mu(n)} (t_n)} \opr{S}_{jk}(t_0)
    \otimes
    \pqty{\prod_{n=1}^{|\mu|} \tilde{E}_{\mu(n)} (t_n)} \tilde{E}_{jk}(t_0)
    .
    \label{eqn:NestedWilcox}
\end{align}
All sequences $\mu$ which include at least one $(j',k')$ with $j'\ne j$ or with any pair $(j',k')$ and $(j'',k'')$ with $j'\ne j''$ represent intra-environment mixing interactions and so must vanish for all possible choices of evaluation times, meaning that we must require
\begin{align}
    \pqty{\prod_{n=1}^{|\mu|} \mathcal{C}_{\mu(n)}(t_n)} \opr{S}_{jk}(t_0) = 0
    \qquad
    \forall j,k,\mu,t_0,t_1,\dots
    ,
    \label{eqn:SopStringZeroTD}
\end{align}
for such sequences.

\subsection{Alternating Qubit Example}
\label{ssec:AlternatingQubit}
%

To illustrate these criteria, we turn to a time-dependent generalization of the qubit model studied in Ref.~\cite{Duruisseau23} and discussed in Sec.~\ref{sec:ExistingWork}.
Specifically, we consider a model of a system qubit interacting with a collection of environment qubits where the system-environment interaction Hamiltonian $\VV_\sn{SE}$ alternates between two otherwise time-independent interactions with some period $\tau$.
In the absence of a free Hamiltonian for either the system or environment, the model is described by the Hamiltonian,
\begin{align}
    \HH(t) = a_\tau(t)\VV_\sn{SE}^a + b_\tau(t)\VV_\sn{SE}^b
    ,
\end{align}
with the time-dependent prefactors
\begin{align}
    a_\tau(t) = 1 - b_\tau(t) \equiv 
    \begin{dcases*}
    1 & if $\lfloor t/\tau\rfloor$ even,
    \\
    0 & if $\lfloor t/\tau\rfloor$ odd.
    \end{dcases*}
\end{align}
If we decompose the interaction Hamiltonians as
\begin{subequations}
\begin{align}
    \VV_\sn{SE}^a &= \sum_{jk}\opr{S}_{jk}^a\otimes\opr{E}_{jk}
    ,
    \\
    \VV_\sn{SE}^b &= \sum_{jk}\opr{S}_{jk}^b\otimes\opr{E}_{jk}
    ,
\end{align}
\end{subequations}
then the double integral of Eq.~\eqref{eqn:FirstTdMixingConstraint} only vanishes if $\comm*{\opr{S}_{j'k'}^b}{\opr{S}_{jk}^a} = 0~~\forall j,j',k,k'$. That is, this commutator must be zero else there is intra-environment mixing incompatible with quantum Darwinism induced by the system-environment interaction.
In fact, given that this is a qubit system the necessity for a time-independent pointer basis to exist also requires that this commutator vanish.
As we will see, violating this criteria does obstruct the emergence of classicality.
 
We simulate three models of this type which vary in their choices of $\opr{S}_{jk}^{a,b}$ and $\opr{E}_{jk}^{a,b}$, two of which we do not expect to exhibit quantum Darwinism and one which we do. 
In each case, the interaction Hamiltonian $\VV_\sn{SE}^{a,b}$ is built from terms of the form $J_i \sigma_0^\alpha\otimes\sigma_{i}^\beta$ for some choice of $\alpha,\beta\in\cbqty{x,y,z}$ and with random coupling coefficients drawn from a unit normal distribution $J_i \sim \mathcal{N}(0,1)$ \footnote{To avoid any potential numerical issues, we modify the definition of $a_\tau$ and $b_\tau$ slightly such that they are not active for the full period over a slightly shrunk interval of $(\tau - 0.01)$.}.
Each model was simulated 500 times with different randomly selected coefficients, where in each simulation there were $N=11$ environment qubits and the joint initial state is a product of eigenstates of $\sigma^y$,
\begin{align}
    \ket{\psi_{y}(0)} &= \frac{\ket{0} + i \ket{1}}{\sqrt{2}}\bigotimes_{k=1}^N \frac{\ket{0_k} + i\ket{1_k}}{\sqrt{2}}
    .
    \label{eqn:QubitAltInitialState}
\end{align}
The time interval between alternations of the interaction Hamiltonian was chosen to be $\tau = 3$ \footnote{This is sufficiently long that some non-negligible information transfer occurs but not so long that any information transfer is complete.}.
 
\subsubsection{Model E: Time-independent interaction}
\begin{figure*}[t]
    \centering
    \includegraphics[width=\textwidth]{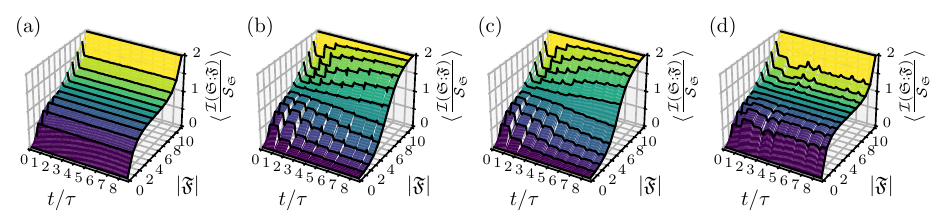}
    \caption{Evolution of the normalized mutual information between the system qubit and environment fragments of a given size $|\sn{F}|\le11$ as a function of time, for (a) model E, (b) model F, (c) model G, (d) model H. We show the average of $500$ randomized choices of coefficients for each Hamiltonian evolving the initial state given in Eq.~\eqref{eqn:QubitAltInitialState}. The white regions on panels (a), (b), and (c) arise because we have constructed these surface plots by drawing the mutual information as a function of fragment size at 100 equally-spaced points in time. Those regions emphasize locations where the variation in the mutual information is large, as there the curves forming the plot become distinguishable and it becomes possible to ``see through the surface''.}
    \label{fig:QubitAlternatingQDExamples}
\end{figure*}

As a point of comparison, we also simulate a time-independent model of a qubit system interacting with a collection of environment qubits described by the Hamiltonian
\begin{align}
    \HH_E = \pqty{\sigz_0 + \sigx_0} \otimes \sum_{i=1}^N J_i \sigz_i
    ,
    \label{eqn:QubitXZHamiltonian}
\end{align}
with random normally distributed couplings $J_i\sim\mathcal{N}(0,1)$.
This model exhibits quantum Darwinism, with the pointer basis being the eigenbasis of $(\sigx+\sigz)$.
The results for this model are shown in Fig.~\ref{fig:QubitAlternatingQDExamples}(a), which plots the evolution of the average mutual information between the system and environment fragments of a given size over time.
For this time-independent model, the plateau characteristic of objectivity emerges extremely quickly on this timescale. 

\subsubsection{Model F: Non-commuting $\opr{S}$, commuting $\opr{E}$}
For our first example with an alternating interaction Hamiltonian, we choose the following Hamiltonian:
\begin{align}
    \HH_F(t) = a_\tau(t)\bqty{\sigz_0\otimes\sum_{i=1}^N J_i \sigz_i} + b_\tau(t)\bqty{\sigx_0\otimes\sum_{i=1}^N K_i \sigz_i}
    ,
\end{align}
with normally distributed coupling constants $J_i, K_i \sim \mathcal{N}(0,1)$.
 
Despite containing all the same terms as the initial time-independent Hamiltonian and at no time having any non-commuting terms simultaneously active, this model does not exhibit quantum Darwinism as demonstrated in Fig.~\ref{fig:QubitAlternatingQDExamples}(b). 
Instead of a plateau in the mutual information plot indicative of redundancy, we see that the mutual information curve is tending to a step at half the size of the environment indicating that small fragments of the environment contain very little system information. 
As discussed, the issue is that there is no time-independent pointer basis in this example since $\sigz$ and $\sigx$ do not commute. 

\subsubsection{Model G: Commuting $V_{\sn{SE}}$}
The next example   is
\begin{align}
    \HH_G(t) = a_\tau(t)\bqty{\sigz_0\otimes\sum_{i=1}^N J_i \sigz_i} + b_\tau(t)\bqty{\sigx_0\otimes\sum_{i=1}^N K_i \sigx_i}
    ,
\end{align}
with $J_i, K_i$ chosen as before.
 
Note that in this case the two operators describing the interaction between the system and any given environment qubit ($\sigz_0\otimes\sigz_i$ and $\sigx_0\otimes\sigx_i$) commute with one another, meaning that this Hamiltonian consists entirely of terms which commute at all times. 
However, as the pointer basis depends only on the system operators this is again a situation where there is no time-independent pointer basis and so no quantum Darwinism as is clear from Fig.~\ref{fig:QubitAlternatingQDExamples}(c), which shows that this model exhibits essentially the same behavior as the previous.

\subsubsection{Model H: Commuting $\opr{S}$, Non-commuting $\opr{E}$}
The only way a time-dependent qubit-qubit model can support quantum Darwinism is if the operators acting on the system in any interaction terms are identical, exactly as in the time-independent case. 
Therefore, for our final example we consider the Hamiltonian
\begin{align}
    \HH_H(t) = \sigz_0\otimes \bqty{a_\tau(t)\sum_{i=1}^N J_i \sigz_i + b_\tau(t)\sum_{i=1}^N K_i \sigx_i}
    ,
\end{align}
with $J_i, K_i$ chosen as before. 
This model satisfies all the requirements necessary to support the emergence of quantum Darwinism, and indeed in Fig.~\ref{fig:QubitAlternatingQDExamples}(d) we do observe a plateau indicating the emergence of objectivity.
 
\begin{figure}
    \centering
    \includegraphics[width=1.0\columnwidth]{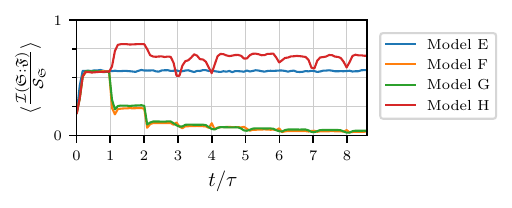}
    \caption{Alternate view of Fig.~\ref{fig:QubitAlternatingQDExamples} showing the average mutual information of fragments with size $|\sn{F}| = 1$ evolving subject to the four example Hamiltonians considered in this section. The mutual information approaches zero for the two models which do not exhibit quantum Darwinism, and whereas it approaches a roughly constant value for those that do.}
    \label{fig:QbQDResD}
\end{figure}

An interesting point to note about this model is that while the system operators always commute with one another -- indicating that there is a time-independent pointer basis which defines the information about the system qubit that can be redundantly encoded into the environment -- the environment operators in the two interaction Hamiltonians do not commute with one another.
This fact is irrelevant as to the binary question of whether or not quantum Darwinism can emerge in this model in the asymptotic limit, but can be very relevant to questions about the process through which it emerges.
This is illustrated in Fig.~\ref{fig:QbQDResD}, which shows the average mutual information between each individual environment qubit and the system as a function of time. 
We observe the alternation between non-commuting environment operators results in each environment qubit containing more information about the system than in the time-independent case.
In the alternating case, the classical plateau is ``sharper'' than in the time-independent case (also visible comparing Fig.~\ref{fig:QubitAlternatingQDExamples}(a) to Fig.~\ref{fig:QubitAlternatingQDExamples}(d)); the mutual information approaches the accessible information more rapidly as a function of the fragment size.
Both cases asymptotically exhibit quantum Darwinism, they differ with regards to the implied constant in the statement that ``a sufficiently large fragment $\sn{F}$ contains all the accessible information about the system''.
 
This example serves to demonstrate an important point: while quantum Darwinism is not possible in the presence of scrambling dynamics in the environment \cite{Duruisseau23}, it is \emph{not} necessary that the environment dynamics be trivial, nor is it necessarily desirable. 
Non-trivial dynamics -- either from the interaction or non-trivial local Hamiltonians -- may lead to enhancements of the emergence of objectivity either in fragment size as in the present case or in the associated timescale. 
They may also present obstructions to objectivity, which may require that a very large environment be studied before objectivity is observed or ensure that the plateau requires a very long time to emerge. 
This is true in both the time-dependent and time-independent cases. 

\section{Collision models}
\label{sec:RepeatedContactModels}
%

\begin{figure}
    \centering
    \includegraphics[width=\columnwidth]{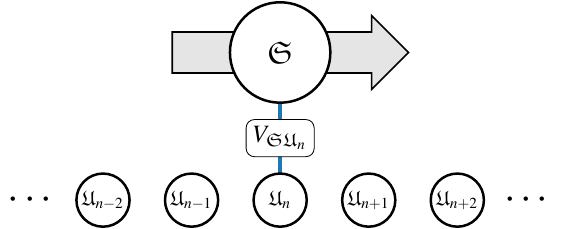}
    \caption{Schematic representation of a collision model with regular and equally-spaced interactions. We imagine the system $\sn{S}$ as flying by a sequence of environment units $\sn{U}$, interacting with each in via some interaction Hamiltonian $\VV_{\sn{SU}_n}$ for a short time.}
    \label{fig:RCMSchematic}
\end{figure}

A particularly general and useful class of time dependent models are the collision models \cite{Ziman05, Strasberg17, Gneiting21, Ciccarello22, Chisholm21, Campbell21}. 
Here, a series of non-interacting environment \emph{units} interact sequentially with (interpreted as a collision with) some system of interest.
These interactions may occur deterministically at regular intervals as illustrated in Fig.~\ref{fig:RCMSchematic}, or stochastically at random times.
 
A variety of important physical models are equivalent to a collision model, for example quantum optical systems involving a stream of atoms interacting with a cavity mode such as the micromaser \cite{Filipowicz86, Rempe90} or models of lasing without inversion \cite{Strasberg17,Scully89,ScullyZubairy}.
In a more general context, collision models can be used as a tool to understand or engineer dynamics. 
For example, it is possible to induce controllable time-dependent Hamiltonian dynamics \cite{Strasberg17} with a collision model by tuning the initial states of the units and their interaction with the system. 
They may also serve as convenient microscopic models for quantum measurements or open quantum dynamics \cite{Strasberg17,Caves87,Karevski09}.
Even more broadly, collision models are intimately connected with and provide concrete framework to study quantum thermodynamics, for example in models related to Maxwell's demon \cite{Mandal12,Mandal13,Deffner13a,Strasberg14,Barato13,Barato14,Chapman15,Zurek2018PR,Safranek2018PRA,Engelhardt2018NJP,Stevens19,Abah2020JPCO,Bhattacharyya2022PRE,Korkmaz2023PRA} or in certain models of quantum heat engines \cite{Scully03}. 
The sequence of units interacting with the system can play the role of a reservoir \cite{Strasberg17} for heat, work, or information \cite{Deffner13b}.
 
The wide applicability of collision models provides a strong motivation to classifying them according to their ability to support quantum Darwinism, and indeed there have been prior examinations of collisional models in this context, e.g. with qubit models \cite{Campbell19}. 
Having such a classification for generic collision models would mean that for physical models which are directly equivalent to a collision model, it would be immediately clear if and how objectivity could emerge from measurements of the stream of units.
For scenarios that can be engineered as effectively coarse-grained collision models, one might hope that an understanding of the requirements necessary for objectivity in the underlying collision model could be translated into insight into the effective model.
And finally, as collision models provide a convenient and flexible example of a general thermodynamic reservoir, it may be possible to build on classification to produce statements about the distinctions between quantum and classical thermodynamics, and the boundary between.
 
A general Hamiltonian describing a collision model is
\begin{align}
    \HH = 
        \Sop{\HH_\sn{S}} 
        + \Eop{\sum_{i=1}^N\HH_{\sn{U}_i}}
        + \sum_{i=1}^N \VV_{\sn{SU}_i}(t)
    ,
    \label{eqn:GenericRCM}
\end{align}
where $\HH_{\sn{U}_i}$ is the free Hamiltonian for the $i$th  unit, and $\VV_{\sn{SU}_i}(t)$ is the interaction Hamiltonian describing the interaction between the system and the $i$th unit,
\begin{align}
    \VV_{\sn{SU}_i}(t) \equiv \theta_i(t) \sum_{k=1}^{|\sn{U}_i|}\opr{S}_{ik}(t)\otimes\opr{E}_{ik}(t) \equiv \theta_i(t) \VV_{\sn{SU}_i}
    ,
\end{align}
where $\theta_n(t)$ is one on some time interval when this particular unit interacts with the system and zero otherwise. 
For example, one common case is where the interactions are regular and equally spaced, where each unit interacts with the system for a time $\delta$ with a periodicity $\tau$,
\begin{align}
    \theta_n(t) = 
    \begin{dcases*}
    1 & if $(n-1)\tau \le t < (n-1)\tau + \delta$,
    \\
    0 & otherwise.
    \end{dcases*}
\end{align}
An arrangement of this type is depicted schematically in Fig.~\ref{fig:RCMSchematic}, where the system is imagined to be moving past a sequence of environment units at some fixed velocity, interacting with each in turn.
Alternately, one could consider random \cite{Chisholm21,GarciaPerez20} or otherwise variable interaction intervals. 
 
As these collision models are just time-dependent system-environment models with a particular structure, our results from the previous section may be directly applied to state the criteria for quantum Darwinism in a collision model:
\begin{itemize}
    \item There must exist a time-independent pointer observable $\opr{\mathcal{A}}=\Sop{\opr{A}}$ which commutes with the Hamiltonian $\HH(t)$ at all times.
    \item The system operators entering the interaction must mutually commute $\comm*{\opr{S}_{ik}(t_1)}{\opr{S}_{i'k'}(t_2)} = 0$ at all times.
    \item If the system Hamiltonian does not commute with the interaction operators then any arbitrarily nested commutator of the free Hamiltonian and interaction operators corresponding to different units must commute with at all times.
\end{itemize}
Since each unit only interacts with the system for a finite interval $\tau$, any information transfer is necessarily irreversible. 
There is no requirement that the coefficients representing the strengths of the system-unit interactions have continuous support, nor that they be random at all so long as they are not chosen such that the unit state is periodic with a period coinciding with the interaction interval $\tau$.
 
Note that if the forms of the system-unit interactions $\VV_{\sn{SU}_i}$ are not fixed it is possible to have non-commuting interactions and still observe quantum Darwinism. 
This is possible if there is a finite prefix of units with possibly non-commuting interactions, with all subsequent units satisfying the criteria for Darwinism.
For example, suppose each unit interacts with the system one after another each for a time $\tau$.
The state of the system qubit after interacting with the prefix of $N_p$ units is
\begin{align}
    \rho_\sn{S}(N_p\tau) = \Tr_{\sn{U}_1,\dots,\sn{U}_{N_p}}\bqty{
        \opr{U}_{N_p}(\tau)\cdots\opr{U}_{1}(\tau)
        \ket{\Psi_{N_p}(0)}
    }
    ,
\end{align}
where $\ket{\Psi_{N_p}(0)} = \ket{\psi_\sn{S}(0)}\bigotimes_{i=1}^{N_p}\ket{\psi_i(0)}$ is the initial state of the system and first $N_p$ units.
It is then the information about \emph{this} new state $\rho_\sn{S}(N_p \tau)$ which is redundantly encoded into the remaining environment units, in the pointer basis defined by the remaining interactions.
Therefore, we choose to interpret such collision models where a finite prefix of non-commuting interactions as a concatenation of a state preparation process followed by an evolution leading to quantum Darwinism.
This is exactly the same reasoning and interpretation we stated for general time-dependent models with a similar cutoff time in the previous section.

We may also consider the case where a stream of units which would otherwise lead to quantum Darwinism are interrupted by a single unit with a non-commuting interaction acting as a perturbation.
Additional interruptions could occur, and would be analyzed similarly.
Before the perturbation, quantum Darwinism emerges. As in other cases, the timescale for this may depend on the initial state if the interactions are asymmetric in how quickly subspaces of the pointer observable dephase.
Then, the non-commuting interaction with the perturbing unit transforms the system state in the pointer basis. 
After the perturbation, the interaction with the remaining units will again lead to quantum Darwinism, but now with the information about the transformed state encoded into the environment units. 
If observers are aware of which units interacted before and after the perturbation, they can choose to measure only units from after the perturbation and hence this scenario is exactly analogous to the initial emergence of quantum Darwinism. 
If observers instead build their environment fragments from a mixture units which interacted before and after the perturbation, then quantum Darwinism may require additional time to emerge as the system information is ``diluted'' by the pre-perturbation units which carry reduced (or even zero) information about the system state depending on the magnitude of the perturbation. 
 
The preceding arguments do not extend to collision models where there is no cutoff beyond which all units interact with the system through commuting interactions, as then there is no point at which a time-independent pointer basis can be defined.
 
The remainder of this section is dedicated to a discussion of a wide range of collision models and the existence or non-existence of quantum Darwinism in each. 
We will first consider a set of simple and illuminating qubit examples which illustrate the points made in our general discussion, then a variety of collision models which have been presented in prior works as useful models for a variety of physical systems, from a quantum Maxwell demon model \cite{Deffner13a} to time-dependent Hamiltonian engineering \cite{Strasberg17}. 

\subsection{Qubit Examples}
\label{ssec:QBRCM}
%

We consider a model of a single system qubit interacting with a series of qubit units, where in each interval the interaction is time-independent,
\begin{align}
    \HH(t) = \sum_{n=1}^N \theta_n(t) J_n \opr{S}_n \otimes \opr{E}_n
    ,
\end{align}
and where we will fix the coefficients all unity, $J_n = 1$.
 
We numerically study four models which differ in the choices made for the operators $\{(\opr{S}_n, \opr{E}_n)\}$ and in how they do or do not satisfy the requirements necessary for quantum Darwinism.
We set the interaction period and interval to $\tau = 1$ and $\delta = 0.95$, respectively, and simulate the dynamics as $N=12$ units starting from the initial state,
\begin{align}
    \ket{\psi_{y}(0)} = \frac{\ket{0} + i\ket{1}}{\sqrt{2}}\bigotimes_{n=1}^N\frac{\ket{0_n} + i\ket{1_n}}{\sqrt{2}}
\end{align}
 
In our analysis, we define the environment to be the set of qubits which have interacted with the system qubit for any interval. For instance, in the first interval $(0,\delta)$ the environment consists of a single qubit ($|\mathfrak{E}|=1$), in the second interval $(\tau,\tau+\delta)$ two qubits, and so on. 
If we imagine an observer measuring fragments of the environment and hoping to observe quantum Darwinism, we are assuming that the observer knows which environment units have interacted with the system and which have not.
This is a reasonable description of a wide range of situations, e.g. each unit might pass through some interaction volume within which it is in contact with the system; the observer then may choose to only measure units which are known to have passed through the interaction volume.
 
If the observer does not have this knowledge and measures fragments that may include environment units which have not yet been in contact with the system, the mutual information between the fragment and the system  grows more slowly with the fragment size due to the inclusion of yet-to-interact units. 
In the long-time limit where the fraction of units which have interacted with the system becomes sufficiently large, these two pictures coincide.
 
\begin{figure*}[t]
    \centering
    \includegraphics[width=\textwidth]{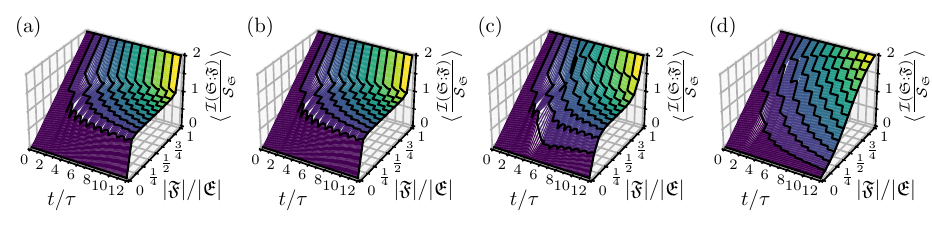}
    \caption{Plots of the normalized mutual information as a function of environment fragment size over time for a the qubit-qubit collision (a) model I, (b) model J, (c) model K, (d) model L. Each figure is the average of 128 simulations, each with a different selection of coupling strengths for the Hamiltonians.}
    \label{fig:QubitRCMQDExamples}
\end{figure*}

We present the results of simulations of these four models in Fig.~\ref{fig:QubitRCMQDExamples}, plotting the average mutual information between fragments of the environment of a given size and the system as a function of time.
Figure~\ref{fig:RCMQDResD} shows the same evolution for a fixed environment fragment size of one unit, which helps visually clarify the difference between the first three cases.

\subsubsection{Model I: Commuting interactions}

The first case   is the simplest, wherein we choose every system-unit interaction $S_n\otimes E_n$ to be of the form $\sigz_0\otimes\sigz_n$.
A very similar model was shown to exhibit quantum Darwinism in Ref.~\cite{Ryan22}, wherein the sequence of units interacting with the system qubit was taken to be an ``inaccessible'' environment, unavailable for measurement. 
The system was coupled to a second accessible environment which could be measured (with the same form of interaction). 
In this setup, objectivity can be observed through measurements of fragments of the accessible environment even as the inaccessible units carry off information about the system.
 
Here, we instead take the units post-interaction to be the environment degrees of freedom accessible to measurement.
This model easily satisfies all the requirements for quantum Darwinism, as evidenced by the clear emergence of a plateau exhibited by our simulations in Fig.~\ref{fig:QubitRCMQDExamples}(a).
The details of the dynamics can be understood from the blue curve in Fig.~\ref{fig:RCMQDResD}, which plots the average mutual information between a single environment unit and the system over time. 
At the beginning of each interval, a new unit begins interacting with the system and so is introduced into the accessible environment. 
This unit has no information about the system initially, hence the average drops immediately upon its introduction. 
As it interacts with the system, information about the system in the $\sigz$ basis is encoded in the unit and so the average grows to a maximum when the unit is decoupled from the system.
 
With finitely many units, this model is essentially equivalent to the qubit model from Ref.~\cite{Duruisseau23} which was discussed in Sec.~\ref{sec:ExistingWork}, where a single system qubit interacts continuously with a bath of environment qubits with $\sigz_0\otimes\sigz_n$ interactions.
The state of that continuous model of one qubit interacting with a bath of size $N$ at time $\delta$ is exactly the same as the state of our collision model after time $N\tau$, i.e., after $N$ units have individually interacted with the system for a time $\delta$ each.
 
This model is also very similar to the stochastic collision model considered in Ref.~\cite{Chisholm22}, which was shown to exhibit quantum Darwinism experimentally using IBM's superconducting quantum hardware.
In that work, the environment units are assumed to interact instantaneously with the system at times distributed according to a Poisson process.
At long times, when the probability that all units have interacted with the system goes to one, the system-environment state of the stochastic model (with a suitably chosen collision strength) approach the same asymptotic state as the collision model, since all the interaction Hamiltonians commute. 

\subsubsection{Models J and K: Prefix of non-commuting interactions}

\begin{figure}[t]
    \centering
    \includegraphics[width=1.0\columnwidth]{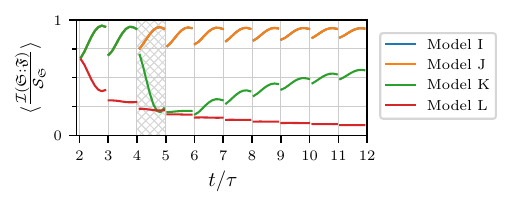}
    \caption{Alternate view of Fig.~\ref{fig:QubitRCMQDExamples} showing the average mutual information of fragments with size $|\sn{F}| = 1$. The plots for models I and J are essentially indistinguishable on this scale, as is that of model K initially. In the interval $\pqty{4\tau, 4\tau+\delta}$, the mutual information for model K drops as the non-commuting interaction perturbs the initial state. Following this, the mutual information slowly rises as information about the new state is encoded into the remaining environment units.}
    \label{fig:RCMQDResD}
\end{figure}

For our second and third examples, we replace the interaction between a single unit and the system in the previous model with $\sigx_0\otimes\sigz_n$. By replacing the $\sigz_0\to\sigx_0$, this new interaction fails to commute with the other $N-1$ system-unit interactions. 
 
Both models have a suffix of commuting interaction terms of the form $\sigz_0\otimes\sigz_n$, however, so we do expect to see the emergence of objectivity eventually.
Numerical results for one of these modifications is shown in Fig.~\ref{fig:QubitRCMQDExamples}(b), where the first $(n=1)$ unit interacts with the system with this new operator.
As we can see, there is little difference visible between this simulation and the previous example; a plateau is still clearly emerging.
To reiterate, this is because the interaction with the first unit perturbed the initial system qubit state $\ket{\psi(0)}\to\opr{\rho}(\tau)$,
at which point the interactions with the remaining units lead to redundant encoding of the information about this new state $\opr{\rho}(\tau)$ in those remaining units. 
The first unit acquired some information about the original qubit state in the $\sigx$ basis, what becomes objective is the information about this perturbed state in the $\sigz$ basis.
 
Figure~\ref{fig:QubitRCMQDExamples}(c) shows the numerical results for another modification of our initial model, now with only the fifth system-unit interaction modified.
The interactions between the system and first four units commute, hence we can clearly see how the emergence of a plateau indicating classicality after the system interacts with these first few units.
Subsequently, the non-commuting interaction with the fifth unit modifies the system state and destroys the plateau. 
We then see a second plateau emerge, indicating that information about this new system state is being redundantly encoded in the remaining units. 
This behavior is more clearly visible in Fig.~\ref{fig:RCMQDResD}, where the green plot corresponding to this scenario matches the evolution in the simple case with no replacement up to the interval starting at $4\tau$. 
Then, under the action of the modified system-unit interaction term, the system qubit state is modified to $\rho(\tau)$, producing the dip in the plot. Finally, all remaining system-unit interactions again work to make the information about this new state in the $\sigz$ basis objective, leading to a slow rise as a second plateau emerges.
 
Essentially, the environment units are partitioned into three classes: the unit that interacted with a non-commuting operator and sets a cutoff, and those units that interacted either before or after the cutoff. 
The final class is the most relevant, as they carry information about the state of the system in its current state and are mostly responsible for the emergence of the plateau.
The information encoded in the first few units is less accurate after the system was perturbed, hence the mutual information between those units and the system is reduced and hence the plateau is washed out.
The degree to which it is reduced depends on the initial state and the magnitude of the perturbation, which is moderate in this case. 
This partitioning is why the second plateau at $t=12\tau$ in Fig.~\ref{fig:QubitRCMQDExamples}(c) is not as broad as the plateaus in Fig.~\ref{fig:QubitRCMQDExamples}(a) or Fig.~\ref{fig:QubitRCMQDExamples}(b), as information about the final system state is only encoded into a subset of the units.
 
Before concluding with this example, it is worth noting that this behavior is not the same as we would find in the case of a continuous interaction between the system and environment qubits with one non-commuting interaction.
There, it is not possible to understand the effect of the non-commuting interaction simply as a redefinition of the system state and so its effect cannot be restricted in the same way, and so a pointer basis cannot be defined.

\subsubsection{Model L: Alternating non-commuting interactions}

Our final example has the units alternate between two different interaction forms, with all odd-numbered units interact with the system according to $\sigz_0\otimes\sigz_{2k-1}$ and all even-numbered units with $\sigx_0\otimes\sigz_{2k}$.
Simulations of this model are shown in Fig.~\ref{fig:QubitRCMQDExamples}(d), where it is clear that no plateau is emerging -- in fact, the mutual information plot is steepening.
This shape is indicative of a lack of redundancy \cite{Riedel12}, implying that determining anything about the system state would require measuring at least half of the total collection of environment units.
 
Unlike the previous two cases, in this example it is not simply one unit interacting with the system in a non-commuting way but instead the units alternate between two non-commuting interaction Hamiltonians.
There is no initial prefix of units which perturb the state after which the remaining units interact in such a way that the information about perturbed state becomes objective as in the previous case.
Hence, no plateau and no quantum Darwinism.
 
This example is more analogous to the continuous model with non-commuting interactions.
There is no cutoff time after which all system-unit interactions agree on a pointer basis, the system state is not stationary, and there can be no objectivity.
 
A similar model where the units act as an inaccessible environment to a system qubit coupled to a second accessible environment was considered in Ref.~\cite{Ryan22}.
There, each unit interacts with the system with the same interaction Hamiltonian, however this interaction fails to commute with the system-accessible environment interaction Hamiltonian. 
The details differ but the conclusion is essentially the same as the example   here, namely that quantum Darwinism cannot emerge in such scenarios.

\subsection{Quantum Maxwell Demon Example}
\label{ssec:DemonRCM}
%

An interesting example of a classical analog of a collision model is the minimal model of a Maxwell demon presented in Ref.~\cite{Mandal12}, where a three-state demon interacts stochastically with a series of classical random and independent bits. 
The stream of incoming bits acts as an information reservoir, for example allowing the demon to exploit biases in the input distribution to perform work.
In the simplest case, the demon eventually approaches a periodic steady state in which any biases in the incoming bitstream induce a steady-state current where the demon cycles through its three states in an order set by the direction of the bias.
 
An analogous model of a quantum Maxwell demon \cite{Deffner13a} replaces the demon with a quantum three-level system (qutrit) and the classical bitstream with a series of qubit units all prepared in some fixed initial state $\rho_\sn{U}(0)$.
 
If we take the three states of the demon to be $\ket{A}$, $\ket{B}$, and $\ket{C}$ then the Hamiltonian of this model of a quantum demon is a collision model of the form shown in Eq.~\eqref{eqn:GenericRCM} with the system Hamiltonian
\begin{align}    
    \HH_\sn{S} &= \dyad{A}{B} + \dyad{B}{A} + \dyad{B}{C} + \dyad{C}{B}
    ,
    \label{eqn:DemonHS}
\end{align}
and system-unit interaction Hamiltonian
\begin{align}
    \VV_{\sn{SU}_n} &= \gamma\big(\dyad{A,1}{C,0} + \dyad{C,0}{A,1}\big)
    ,
    \label{eqn:DemonV}
\end{align}
where we use the notation $\ket{A,1} \equiv \ket{A}\otimes\ket{1}$. 
The behavior of this model is similar to the classical demon model \cite{Deffner13a}, in that the demon qutrit eventually approaches a periodic steady state which may have some persistent current $\ket{A}\to\ket{B}\to\ket{C}\to\ket{A}$ of either sign depending on the state in which each qubit unit is prepared.
 
This quantum demon model is interesting to consider from the point of view of quantum Darwinism because it is such a straightforward generalization of a classical model, and so one might imagine that there could be some sense in which measuring the quantum model in some way might lead to a reduction to the classical model coinciding with the emergence of quantum Darwinism.
 
This cannot be the case, however.
Intuitively, this is due to the fact that the interaction between the demon and the sequence of units induces a steady-state current circulating within the full three-dimensional state space of the demon.
There is no pointer basis in which the demon eventually dephases, so there is no quantum Darwinism.
This can be easily verified by checking that there is no pointer observable which commutes with both the system Hamiltonian and system-unit interaction.
 
\begin{figure}
    \centering
    \includegraphics{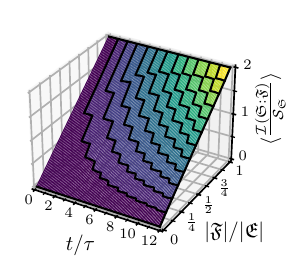}
    \caption{Evolution of the mutual information between a three-level system and $|\sn{E}|=12$ environment units (qubits) subject to the Hamiltonians of Eqs.~\eqref{eqn:DemonHS} and \eqref{eqn:DemonV}. No plateau emerges, and instead the mutual information remains linear in the fragment size indicating that there is no redundant information encoded in the environment.}
    \label{fig:DemonQD}
\end{figure}

Figure~\ref{fig:DemonQD} shows the results a simulation of this quantum demon model with 12 environment qubits with interaction strength $\gamma=(4/3)$, where the interaction period was $\tau = 1$ and each unit interacted for a time $\delta = 0.95$.
The initial state of the demon was $\ket{\psi_\sn{S}(0)} \propto \ket{A}+\ket{B}+2i\ket{C}$ and each unit was initialized in $\ket{\psi_\sn{U}(0)} \propto \ket{0} + 2i\ket{1}$.
As in our previous simulations, the environment fragments are drawn from units after interacting with the demon.
We see that the mutual information is linear in the size of the fragment at all times, indicating that there is no redundant encoding of information about the demon in the environment units whatsoever.

These results raise several interesting questions: 
While we have shown that this particular model of a quantum Maxwell demon does not exhibit quantum Darwinism in any parameter regime, we have not shown that such a model cannot exist. 
Further, it is possible that demanding a demon models exhibit quantum Darwinism asymptotically is overly restrictive when attempting to connect to classical models. 
It may be that a precise formulation of what such a correspondence should entail would instead point to some broader concept of objectivity of thermodynamic properties, for example though some variation of the predictability sieve idea \cite{Zurek93}. 
The present classification will inform future work in these directions, both in the design of models and in helping show which assumptions may be unwarranted or unnecessary. 

\subsection{Other Collision Models}
\label{ssec:OtherRCMs}
%

Based on the criteria we have presented in this work, we may draw some fairly general conclusions about the types of collision models which could be expected to exhibit quantum Darwinism when considering measurements of the units post-interaction.
Most importantly, the emergence of objectivity requires that there be a time-independent pointer observable which defines what information about the system becomes objective. 
Time-independence is crucial, so that the system state remains stable such that the repeated interactions of the system and its environment can redundantly encode this fixed information in the environment. 
 
For many practical use cases of collision models, it is not possible to require that there be some fixed pointer observable that is also compatible with the application the model is designed around.
For example, in the previous section we showed that a minimal model of a quantum Maxwell demon built around a collision model does not exhibit quantum Darwinism.
In essence, the issue is that the system-unit interaction Hamiltonian is structured to drive unit-dependent transitions in the demon system. 
This is fundamentally incompatible with quantum Darwinism, where the system state should remain unperturbed while encoding information into the environment units. 
It may be possible to construct a model of a quantum Maxwell demon which supports quantum Darwinism, however the information about the system state which becomes objective can not be directly related to its function as a model of a quantum Maxwell demon.
 
Another similar example is given by the collisional description of a micromaser \cite{Strasberg17}. 
Here, a series of two-level atom units pass through a microwave resonator and interact via a Jaynes-Cummings interaction,
\begin{subequations}\begin{align}
    \HH_\sn{S} &= \omega \opr{a}^\dagger\opr{a}
    \\
    \HH_{\sn{U}_n} &= \frac{\delta}{2}\sigz_n
    \\
    \VV_{\sn{SU}_n} &= g\pqty{\opr{\sigma}_n\opr{a}^\dagger + \opr{\sigma}_n^\dagger\opr{a}}
\end{align}\end{subequations}
where $\opr{a}^\dagger,\opr{a}$ are the creation and annihilation operators for the cavity mode and $\opr{\sigma}_n^\dagger,\opr{\sigma}_n$ the raising and lowering operators for the $n$'th atom.
With carefully chosen parameters, it is possible to ensure that the net results of each interaction is that each atom unit transfers energy to the microwave mode, pumping the field.
 
Although the main focus of our discussions has been centered around finite-dimensional systems, it is not difficult to see that this model does not support quantum Darwinism. 
There is no pointer observable which is diagonal in the number basis (and so commutes with $\HH_\sn{S}$) which commutes with the operators $\opr{a}$ and $\opr{a}^\dagger$ entering the interaction. 
Indeed, as in the demon model the system-environment interaction is specifically designed to drive transitions in the system.
 
The same conclusion can be drawn for essentially any other arrangement where the system is a simple harmonic oscillator which interacts with each environment unit according to a Jaynes-Cummings type interaction,
\begin{align}
    \VV_{\sn{SU}_n} \propto \pqty{\opr{a}^\dagger \opr{O} + \opr{a} \opr{O}^\dagger}
    ,
\end{align}
for example swapping the two-level atoms of the micromaser model for three-level atoms to study lasing without inversion \cite{ScullyZubairy, Mompart00, Strasberg17} or related ideas \cite{Scully03}.
No pointer basis can be found in these models, hence they cannot exhibit quantum Darwinism.
 
There do exist some applications of collision models which can support quantum Darwinism, however. 
For example, it is possible \cite{Strasberg17} to engineer time-dependent Hamiltonians acting on the system,
\begin{align}
    \HH_\sn{S}(t) = \HH_\sn{S} + \lambda(t)\opr{O}
    ,
    \label{eqn:EngTDHamiltonian}
\end{align}
by interacting the system with a series of units with the interaction
\begin{align}
    \VV_{\sn{SU}_n} \propto \opr{O}\otimes\opr{E}_n
\end{align}
with the operators $\opr{E}_n$ and initial states for each unit $\rho_{\sn{U}_n}(0)$ specifically chosen such that the effective dynamics reproduce Eq.~\eqref{eqn:EngTDHamiltonian}.
 
Here, whether or not quantum Darwinism emerges, depends on the relationship of the engineered time-dependent Hamiltonian and the free system Hamiltonian. 
If there exists a pointer observable that commutes with both, one should expect to observe the information about the system in the associated basis become objective in measuring the units post-interaction.
In this case, the time-dependent Hamiltonian corresponds to shifting the rates at which phases are accumulated in this bases -- phases which are irrelevant to the system-environment interaction.

\section{Concluding remarks}
\label{sec:Conclusion}
%

In this work, we have presented a classification of a large class of generic system-environment models incorporating two-body interactions which may or may not be time-dependent according to whether they support the emergence of classical objectivity.
 
The necessity of a pointer basis (defined by a pointer observable) which specifies what information about the system becomes objective places simple constraints on the commutation relations between the pointer observable and the system operators entering the Hamiltonian. 
Hamiltonians satisfying these constraints are guaranteed to preserve the pointer states, and provide a generalization of the separable ``parallel decoherent interaction'' Hamiltonians for qubit systems introduced in Ref.~\cite{Duruisseau23}.
 
Quantum Darwinism and the emergence of objectivity require additionally that the joint system-environment Hamiltonian preserves states of singly-branching form, which are those states representing the redundant information encoding emblematic for quantum Darwinism \cite{Touil22}.
This is the case if the Hamiltonian does not induce any information scrambling in the environment \cite{Duruisseau23}, which would disperse the system information encoded into the environment in such a way that it can no longer be recovered by measurements of environment fragments.

In the generic context of the present work, this requirement translates to another set of constraints on commutation relations between the system operators entering the Hamiltonian.
Essentially, we may summarize these constraints intuitively by saying that the interaction between the system and the environment degrees of freedom must not induce any effective mixing or scrambling dynamics in the environment. 
Note that this statement is in a certain sense dependent on how exactly one distributes individual physical degrees of freedom in the environment into fragments to measure. 
As we have shown, mixing on disjoint subsets of the environment is not necessarily incompatible with quantum Darwinism.
 
The model we have taken as the basis for our classification presented in this work is general enough to apply to a number of relevant and realistic scenarios.
Crucially, our results are sufficiently general that they may be used to understand the emergence of classicality in collision models as discussed in Sec.~\ref{sec:RepeatedContactModels}. 
Such models may be used to analyze a huge range of applications, most obviously those which are manifestly equivalent to a collision model such as the minimal model of a quantum Maxwell demon \cite{Deffner13a} or the micromaser \cite{Filipowicz86}, both examples which were analyzed using our results in Sec.~\ref{sec:RepeatedContactModels}.
More importantly, it can be shown \cite{Strasberg17} that collision models can be used to effectively act as thermodynamic reservoirs, generators of open quantum dynamics, generators of time-dependent Hamiltonian, and more.
Based on our understanding of quantum Darwinism in collision models, it should therefore be possible to bootstrap an understanding of quantum Darwinism and to study the boundary between quantum and classical behavior in these broader contexts.
 
Following our results presented here, there are clear paths for further generalizations in future work.
For example, one could consider more general classes of degrees of freedom for the system or environment, e.g. discrete infinite-dimensional or continuous-variable subsystems. 
The structure of the arguments we have used to arrive at our classification should apply in these more general contexts, with differences in the details due to the structure of operators acting on these more complicated systems. 
Alternatively, one could allow the Hamiltonian to include three- or more body interaction terms which cannot be interpreted as two-body interactions either between the system and a bipartite environment degree of freedom or between a composite system and a single environment degree of freedom.
Intuitively, three-body interactions appear to be at odds with the ``no environment mixing'' requirement, however properly treating this case will likely be a much larger task than the two-body case.

\acknowledgements{S.D. acknowledges support from the John Templeton Foundation under Grant No. 62422.}

\appendix
%

\section{Constraints in the time-independent case}
\label{app:DeriveHamiltoniansPointerBasis}
This appendix presents a detailed justification for the criteria for a generic time-independent Hamiltonian to exhibit quantum Darwinism presented in Sec.~\ref{sec:Model}.

\subsection{Existence of a pointer basis}
The generic Hamiltonian presented in Eq.~\eqref{eqn:GenericHamiltonian} supports a pointer basis if and only if there exists some pointer observable $\opr{\mathcal{A}} \equiv \Sop{\opr{A}}$ with which it commutes. 
This pointer observable must commute with the system Hamiltonian ($\comm*{\opr{A}}{\HH_\sn{S}} = 0$), and it must also commute with the system-environment interaction.
 
This commutator is given by,
\begin{align}
    \comm{\opr{\mathcal{A}}}{\HH_\sn{SE}} 
    &= \comm{\Sop{\opr{A}}}{\sum_{i=1}^m \opr{S}_i\otimes\opr{E}_i}
    \\
    &= \sum_{i=1}^m \comm{\opr{A}}{\opr{S}_i}\otimes\opr{E}_i
    \\
    &= \sum_{j=1}^{|\sn{E}|}\sum_{k=1}^{|\mathcal{L}(\sn{E}_j)|}\sum_{l=1}^{N_S(j)} \comm{\opr{A}}{\opr{S}_{jkl}}\otimes\opr{E}_{jk}
\end{align}
wherein we have reinterpreted $i$ as a multi-index and rewritten the sum over system-environment interaction terms as a sum over environment sites (indexed by $j$), over some basis for the operators acting on that site (indexed by $k$ from $1$ to $|\mathcal{L}(\sn{E}_j)|$), and over the different system operators which are coupled to each of these operators (indexed by $k$).
 
Each term in the double sum over $j$ and $k$ corresponds to a linearly independent environment operator, therefore for the overall sum to be zero each term must vanish individually.
For this to happen, we must require that
\begin{align}
    \comm{\opr{A}}{\sum_{l=1}^{N_S(j,k)}\opr{S}_{jkl}} = 0 \qquad \forall j,k
    .
\end{align}
In other words, it must be possible to collect the various $\opr{S}_{jkl}$ into a set of $\opr{S}_{jk}'$ which all commute with the pointer observable, 
\begin{align}
    \comm{\opr{A}}{\opr{S}_{jk}'} = 0
    ,
\end{align}
in which case the interaction Hamiltonian can be written as
\begin{align}
    \sum_{j=1}^{|\sn{E}|}\sum_{k=1}^{|\mathcal{L}(\sn{E}_j)|}\opr{S}_{jk}' \otimes \opr{E}_{jk}
    \label{eqn:APPPointerGoodInteractionHamiltonian}
\end{align}
We drop the prime through the remainder of this appendix and paper, and assume that the operators acting on the system have been combined in this way.

\subsection{Non-commuting system operators}
If the pointer observable is non-degenerate, then it is necessarily true that any two system operators $\opr{S}_{jk}$ and $\opr{S}_{j'k'}$ which commute with $\opr{A}$ and so preserve the pointer basis commute with one another.
However, if the pointer observable is degenerate (as is possible with a system larger than a qubit), then it is possible for $\HH$ to support a pointer ``basis'' even while $\comm{S_{jk}}{S_{j'k'}} \ne 0$. 
The resulting dynamics preserves the projections of the initial state onto each degenerate subspace of $\opr{A}$, each of which effectively plays the role of a pointer state.
 
Such Hamiltonians do not support quantum Darwinism, however, as they fail to preserve the necessary singly-branching structure of the composite system-environment state. 
To show this, we consider a simple system-environment interaction Hamiltonian
\begin{align}
    \HH = \opr{S}_1\otimes\opr{E}_1 + \opr{S}_2\otimes\opr{E}_2
    ,
\end{align}
where $\opr{E}_{1,2}$ are assumed to be operators acting on distinct environment degrees of freedom. 
If $\comm{\opr{S}_1}{\opr{S}_2} = 0$, then the propagator
\begin{align}
    \opr{U}(t) = e^{-it\pqty*{\opr{S}_1\otimes\opr{E}_1 + \opr{S}_2\otimes\opr{E}_2}}
    ,
\end{align}
factors trivially into two propagators which can easily be seen to preserve the singly-branching form.
If instead the system operators do not commute, we can express the propagator as an infinite product using the Zassenhaus \cite{Casas12} formula
\begin{align}
    \opr{U}(t) = 
        e^{-it\pqty*{\opr{S}_1\otimes\opr{E}_1}}
        e^{-it\pqty*{\opr{S}_2\otimes\opr{E}_2}}
        e^{\frac{t^2}{2}\comm{\opr{S}_1}{\opr{S}_2}\otimes\opr{E}_1\otimes\opr{E}_2}
        \dots
    \label{eqn:APPPropagatorZassenhaus}
\end{align}
The third factor in this product (and higher-order factors not shown here) represent an induced interaction between the two different environment sites, mediated by the system due to the nonvanishing commutator $\comm{\opr{S}_1}{\opr{S}_2}$.
This mixing is incompatible with the singly-branching form, hence we conclude that quantum Darwinism does not arise in models with interaction Hamiltonians of the form shown in Eq.~\eqref{eqn:APPPointerGoodInteractionHamiltonian} if any pair of system operators  fail to commute, provided they correspond to interactions with distinct environment degrees of freedom.
 
Note that the singly-branching form would have been preserved in this simple example if we had taken the two environment operators to act on the same site,
\begin{align}
    \HH = \opr{S}_1\otimes\opr{E} + \opr{S}_2\otimes\opr{E}'
    ,
    \label{eqn:APPHamOneSiteMixing}
\end{align}
as then the factors in Eq.~\eqref{eqn:APPPropagatorZassenhaus} and the higher-order factors do not lead to a violation of any of the requirements for quantum Darwinism: since $\comm*{A}{S_1} = \comm*{A}{S_2} = 0$, nested commutators of $S_1$ and $S_2$ also commute with $A$ and therefore support the same pointer basis. Additionally, since $\opr{E}$ and $\opr{E}'$ act on the same environment degree of freedom, the singly branching form is preserved by each term in the infinite product.
 
This is not a particularly interesting conclusion in the case when the system is finite-dimensional, as then only a finite number of environment degrees of freedom could interact with the system through non-commuting operators (limited by the dimensionality of the space of operators acting on the system and the need for the remainder of the environment to interact with commuting operators).
When taking the limit of a large environment, it then makes more sense to consider this finite subset of degrees of freedom as either a second auxiliary environment or as part of the system itself. 
For this reason, we do not consider this possibilty in the present work.
 
More usefully, a very similar argument to that just presented shows that while the system-environment interaction may not have non-commuting system operators, it is acceptable to have non-commuting operators acting on any given environment site enter the interaction.
 
For example, the Zassenhaus formula applied to the propagator under the Hamiltonian \eqref{eqn:APPHamOneSiteMixing} would yield
\begin{align}
    \opr{U}(t) = 
        e^{-it\pqty*{\opr{S}_1\otimes\opr{E}}}
        e^{-it\pqty*{\opr{S}_2\otimes\opr{E}'}}
        e^{\frac{t^2}{2}\opr{S}_1\opr{S}_2\otimes\comm{\opr{E}}{\opr{E}'}}
        \dots
    ,
\end{align}
where here we do not assume that $\opr{E}$ and $\opr{E}'$ commute.
Arbitrary polynomials of $\opr{S}_1$ and $\opr{S}_2$ commute with $\opr{A}$ if $\opr{S}_1$ and $\opr{S}_2$ do, therefore each factor in the product supports the same pointer basis. The singly-branching form is preserved by each factor as well, as any arbitrarily deeply nested commutator of $\opr{E}$ and $\opr{E}'$ is merely another operator on the same site, hence there is no obstruction to quantum Darwinism due to non-commuting environment operators. 

\subsection{Non-commuting free Hamiltonians}
We also consider the possibility that in addition to the system-environment interaction, there may be obstacles to the emergence of objectivity arising from free system or environment Hamiltonians which preserve the pointer basis but which do not commute with the interaction.

\subsubsection{System Hamiltonian}
When the pointer observable $\opr{A}$ is degenerate, it is possible that $\HH_\sn{S}$ and $\opr{S}_{jk}$ fail to commute with each other while individually commuting with $\opr{A}$.
In this case, there are additional constraints on the system operators which enter the interaction Hamiltonian which must be satisfied to avoid the introduction of mixing between distinct environment degrees of freedom.
 
Consider the Hamiltonian
\begin{align}
    \HH = \Sop{\HH_\sn{S}} + \opr{S}_1\otimes\opr{E}_1 + \opr{S}_2\otimes\opr{E}_2
    ,
\end{align}
assuming $\comm*{\opr{S}_1}{\opr{S}_2} = 0$ but $\comm*{\HH_\sn{S}}{\opr{S}_j} \ne 0$.
As in our earlier discussion, we can expand the propagator describing time evolution under the action of this Hamiltonian into an infinite product using the Zassenhaus formula, this time to third order, yielding
\begin{widetext}
\begin{align}
    \opr{U}(t) &= 
        e^{-it\pqty*{\Sop{\HH_\sn{S}}}}
        e^{-it\pqty*{\opr{S}_1\otimes\opr{E}_1 + \opr{S}_2\otimes\opr{E}_2}}
        e^{\frac{t^2}{2}\pqty{
            \opr{S}_1'\otimes\opr{E}_1
            +
            \opr{S}_2'\otimes\opr{E}_2
        }}
        \nonumber\\&\qquad\times
        e^{i\frac{t^3}{3}\pqty{
            \comm{\opr{S}_1}{\opr{S}_1'}\otimes\opr{E}_1^2
            +
            \comm{\opr{S}_2}{\opr{S}_2'}\otimes\opr{E}_2^2
            +
            \comm{\opr{S}_1}{\opr{S}_2'}\otimes\opr{E}_1\otimes\opr{E}_2
            +
            \comm{\opr{S}_2}{\opr{S}_1'}\otimes\opr{E}_1\otimes\opr{E}_2
            +
            \frac{1}{2}\comm{\HH_\sn{S}}{\opr{S}_1'}\otimes\opr{E}_1 
            +
            \frac{1}{2}\comm{\HH_\sn{S}}{\opr{S}_2'}\otimes\opr{E}_2
        }}
        \dots
    ,
\end{align}
\end{widetext}
where for brevity we have introduced the notation $\opr{S}_j'$ for the commutator $\comm*{\HH_\sn{S}}{\opr{S}_j}$.
We see that there are contributions to the third-order factor which represent effective mixing interactions between different degrees of freedom in the environment (the terms with $\opr{E}_1\otimes\opr{E}_2$) if the commutators of $\HH_\sn{S}$ with $\opr{S}_{1,2}$ produce new system operators which fail to commute.
 
For example, suppose the pointer observable $\opr{A}$ supports a single degenerate subspace of dimension 3 spanned by the basis vectors $\cbqty{\ket{\tilde{0}}, \ket{\tilde{1}}, \ket{\tilde{2}}}$, and suppose that the system operators act only on this subspace for simplicity.
If we take the following three operators which all commute with $\opr{A}$,
\begin{subequations}\begin{align}
    \HH_\sn{S} &= \dyad*{\tilde{0}}{\tilde{2}} + \dyad*{\tilde{2}}{\tilde{0}}
    ,
    \\
    \opr{S}_1 &= \dyad*{\tilde{0}} - \dyad*{\tilde{1}}
    ,
    \\
    \opr{S}_2 &= \dyad*{\tilde{0}} - \dyad*{\tilde{2}}
    ,
\end{align}\end{subequations}
then the nested commutators
\begin{align}
    \comm{\opr{S}_1}{\comm{\HH_\sn{S}}{\opr{S}_2}} = \comm{\opr{S}_2}{\comm{\HH_\sn{S}}{\opr{S}_1}} 
    =
    -2\HH_\sn{S}
    ,
\end{align}
do not vanish.
Therefore the resulting dynamics would not support quantum Darwinism due to mixing induced through a combination of the system-environment interaction and the system Hamiltonian.
 
Note that since these mixing dynamics are a higher-order effect, we do expect that they only become relevant at longer times, and that at shorter times there may be some transient emergence of objectivity before the mixing dynamics takes over.
In this work we concern ourselves with the asymptotic emergence of objectivity, therefore we reject that these models exhibit Darwinism. 
 
At fourth order in the expansion, commutators of the forms $\comm*{\HH_\sn{S}}{\comm*{\HH_\sn{S}}{\comm*{\HH_\sn{S}}{\opr{S}_j}}}$, $\comm*{\opr{S}_k}{\comm*{\HH_\sn{S}}{\comm*{\HH_\sn{S}}{\opr{S}_j}}}$, and $\comm*{\opr{S}_l}{\comm*{\opr{S}_k}{\comm*{\HH_\sn{S}}{\opr{S}_j}}}$ appear, each with a corresponding environment operator which is a product of the associated $\opr{E}_{j,k,l}$.
To prevent effective mixing dynamics in the environment, we must require that the second and third commutators vanish whenever any of the indices correspond to different environment degrees of freedom.
The same is true at higher orders, with the commutators corresponding to an arbitrary string of $\HH_\sn{S}$ and $\opr{S}_x$ commuted with $\comm*{\HH_\sn{S}}{\opr{S}_j}$.
 
Incorporating all the constraints from all orders in the expansion, to allow quantum Darwinism to emerge in the asymptotic limit we must require that commuting $\comm{\HH_\sn{S}}{\opr{S}_{jk}}$ arbitrarily many times with the system Hamiltonian $\HH_\sn{S}$ or any system operator $\opr{S}_{j'k'}$ must produce zero if the set of corresponding environment operators $\{\opr{E}_{jk},\opr{E}_{j'k'},\dots\}$ contains operators acting on distinct degrees of freedom in the environment.
In the main text, we introduced a superoperator notation which allows this statement to be expressed compactly as an equation [Eq.~\eqref{eqn:SopStringZero}].
 
In practice, we generally simplify these constraints to requiring that all the system operators commute with one another and the free Hamiltonian,
\begin{subequations}
\begin{align}
    \comm{\HH_\sn{S}}{\opr{S}_{jk}} &= 0 \qquad \forall j,k
    \\
    \comm{\opr{S}_{j'k'}}{\opr{S}_{jk}} &= 0 \qquad \forall j\ne j',k,k'
\end{align}
\label{eqn:APPSimpleConstraints}
\end{subequations}

\subsubsection{Environment Hamiltonian}
Unlike the free system Hamiltonian, there are no constraints placed on commutators involving the free Hamiltonian for each environment degree of freedom, $\HH_{\sn{E}_k}$.
The reason is that since we have assumed that the environment degrees of freedom are separate and that the full environment free Hamiltonian is separable, we consider each environment degree of freedom in isolation.
Then, any arbitrary combination of $\HH_{\sn{E}_n}$ and any interaction operators $\opr{E}_{nk}$ is just another operator acting on the same subsystem $\sn{E}_n$.
Since no intra-environment interactions can be induced by the addition of a non-commuting free Hamiltonian for any set of environment sites nor can they affect the preservation of the pointer basis, such free Hamiltonians cannot alter the ability of the joint Hamiltonian to satisfy any of the requirements for quantum Darwinism.
 
It is possible for the details of the internal dynamics of each environment subsystem to affect the emergence of objectivity, however, as they may aid or hinder the ability to encode system information into the environment. 
This may change the timescales on which objectivity emerges or the minimum sizes of environment fragments $|\sn{F}|$ necessary to recover the accessible information about the system, however in the asymptotic limit Darwinism emerges independent of these details. 

\section{Effects of initial correlations}
\label{app:InitialStates}
\begin{figure*}[t]
    \centering
    \includegraphics[width=\textwidth]{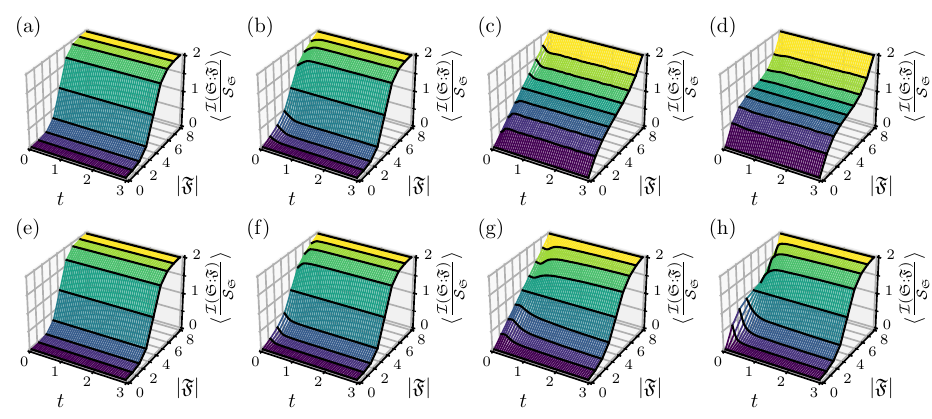}
    \caption{Evolution of the normalized mutual information between the system qutrit and environment fragments of a given size as a function of time for the qutrit-qubit model of Sec.~\ref{sec:Qutrit}. For (a-d), the Model A Hamiltonian [Eq.~\eqref{eqn:QutritHGloballySep}] was used and for (e-h) the Model D Hamiltonian [Eq.~\eqref{eqn:QutritHBad}] was used. Each plot shows the mutual information averaged over $96$ simulations with randomized choices of coefficients for each Hamiltonian. Different forms of initial states were used: (a,e) globally-entangled, (b,f) separable, (c,g) product, and (d,h) singly-branching, as defined in Eqs.~\eqref{eqn:APPInitialStatesAB}, \eqref{eqn:APPInitialStatesC}, and~\eqref{eqn:APPInitialStatesD}. For each type of initial state, one single random example was generated and used to provide the same initial conditions for all simulations shown in this figure.}
    \label{fig:APPQutritDifferentStates}
\end{figure*}
In this appendix, we present numerical results which demonstrate the effects of different initial system-environment states on the emergence of quantum Darwinism.

As discussed in Sec.~\ref{ssec:SinglyBranchingForm}, there is a straightforward and intuitive criteria to understand whether or not an initial state will exhibit objectivity, when propagated forward in time under the action of a Hamiltonian compatible with quantum Darwinism: the state must be of singly-branching form.
If the state is not of this form, meaning it has initial intra-environmental correlations, then these correlations must be removed to allow objectivity to emerge \cite{Touil22}.
We know that Hamiltonians which support quantum Darwinism cannot remove these correlations, however, since such Hamiltonians may not include any direct or effective interactions between environment degrees of freedom. 
Hence, initial environment correlations are incompatible with the asymptotic emergence objectivity. 

To demonstrate this fact we have performed simulations of the model consisting of a single system qutrit and an environment consisting of a collection of qubits from Sec.~\ref{sec:Qutrit}, specifically the models denoted A and D in that section. Model A has a separable interaction Hamiltonian [Eq.~\eqref{eqn:QutritHGloballySep}] which supports quantum Darwinism, and model D has an interaction Hamiltonian [Eq.~\eqref{eqn:QutritHBad}] with non-commuting system operators which does not exhibit quantum Darwinism.

As we are interested in the effects of initial correlations, each of the two models was tested with four different initial states. 
Defining the joint ground state as 
\begin{align}
    \ket{\mathbf{0}_{\sn{SE}}} = \ket{0}\bigotimes_{i=1}^N \ket{0_i}
    ,
\end{align}
three of the four initial states considered were constructed by acting a Haar-random unitary operator on this joint ground state, with the difference being the set of unitary operators considered.
In the first case, the unitary operator $\mathcal{U}_{\sn{SE}}$ was drawn from the space of unitary operators transforming the full joint space. 
The second case built the initial state with two Haar-random unitary operators, one transforming the system only $\mathcal{U}_{\sn{S}}$ and one transforming the full environment space $\mathcal{U}_{\sn{E}}$.
Explicitly, these states are,
\begin{subequations}
\begin{align}
    \ket{\psi_{\rm ent}(0)} &= \mathcal{U}_{\sn{SE}} \ket{\mathbf{0}_{\sn{SE}}}
    ,
    \\
    \ket{\psi_{\rm sep}(0)} &= \pqty{\mathcal{U}_{\sn{S}} \otimes \mathcal{U}_{\sn{E}}} \ket{\mathbf{0}_{\sn{SE}}}
    .
\end{align}
    \label{eqn:APPInitialStatesAB}
\end{subequations}
States built with either of these methods generically include significant correlations between environment degrees of freedom, which cannot be removed by any Hamiltonian which supports quantum Darwinism since such Hamiltonians do not include intra-environmental interactions capable of affecting a change in the correlation structures. This is borne out in panels (a) and (b) of Fig.~\ref{fig:APPQutritDifferentStates}. 
The mutual information plots exhibit the step-like behavior characteristic of ``generic'' system-environment states, indicating that there is no accessible, redundant encoding of system information present.
Note that while each panel in Fig.~\ref{fig:APPQutritDifferentStates} represents an average over Hamiltonians with randomly chosen coefficients, the initial states $\ket{\psi_{\rm ent}(0)}$ and $\ket{\psi_{\rm sep}(0)}$ were generated only once and each trial then used the same initial conditions.

As we have shown, Hamiltonians which support the correlation structure necessary for quantum Darwinism cannot remove initial correlations incompatible with the singly-branching form of the system-environment state.
If the initial state does not contain such correlations, we expect to see the emergence of quantum Darwinism.
The remaining two initial states tested demonstrate this.
The first is a product state, where the system and each environment degree of freedom are prepared with a Haar-random unitary drawn from their respective spaces of local operators,
\begin{align}
    \ket{\psi_{\rm prod}(0)} &= \pqty{\mathcal{U}_{\sn{S}} \bigotimes_{j=1}^{|\sn{E}|} \mathcal{U}_{\sn{E}_i}} \ket{\mathbf{0}_{\sn{SE}}}
    ,
    \label{eqn:APPInitialStatesC}
\end{align}
and the second is a state of singly-branching form. 
The conditional environment states corresponding to the three qutrit levels (i.e., the states on each branch) are each randomly-generated product states, with each branch having equal weight,
\begin{align}
    \ket{\psi_{\rm sbf}(0)} &= \frac{1}{\sqrt{3}}\sum_{n=0}^2 \ket{n} \bigotimes_{i=1}^{|\sn{E}|} \mathcal{U}_{\sn{E}_i}^{(n)} \ket{0_i}
    .
    \label{eqn:APPInitialStatesD}
\end{align}
Since neither of these initial states include correlations between environment degrees of freedom, we find that the action of the Hamiltonian does lead to the emergence of quantum Darwinism as exhibited in panels (c) and (d) of Fig.~\ref{fig:APPQutritDifferentStates}.

While considering the effects different choices of initial state have, it is important to keep in mind that this these effects are always secondary to the choice of Hamiltonian. 
The Hamiltonian determines the dynamics, and the primary consideration is whether or not those dynamics support the emergence of objectivity. 
It is true that highly non-classical initial states can inhibit quantum Darwinism, but the converse is not true.
For example, panels (e-h) of Fig.~\ref{fig:APPQutritDifferentStates} show that all four types of initial state considered do not exhibit a plateau in the mutual information at late times when the Hamiltonian does not support quantum Darwinism. 
Instead, in each case the plot develops step-like features indicating that the dynamics generates generic entangled states. 
This is true even in the case where the initial state is of singly-branching form as shown in panel (h); here we see at \emph{short} times a plateau in the mutual information plot indicating objectivity, which the Hamiltonian quickly destroys. 

Before concluding our discussion of initial state effects it should be noted that -- like many of the constraints presented in this work -- it is possible to weaken the requirement that the initial state have no intra-environment correlations whatsoever.
The logic is exactly the same as that presented in Sec.~\ref{ssec:SinglyBranchingForm} when discussing weakening the requirement that environment degrees of freedom be non-interacting. 
Correlations that are sufficiently weak or sufficiently sparsely distributed will not appreciably affect the emergence of quantum Darwinism, as so long as the initial state is sufficiently close to a state of singly-branching form, objectivity will emerge.
This can be made precise with $\epsilon-\delta$ approaches to quantum Darwinism, however in keeping with the spirit of the classification we present we leave this for later work.

\section{Constraints in the time-dependent case}
\label{app:DeriveTdHamiltonians}
In this appendix, we will generalize the criteria we derived in Appendix~\ref{app:DeriveHamiltoniansPointerBasis} for a Hamiltonian to support quantum Darwinism to the time-dependent case. 
The time-dependent system-environment Hamiltonian is,
\begin{subequations}
\begin{align}
    \HH(t) &= \Sop{\HH_\sn{S}(t)} + \Eop{\HH_\sn{E}(t)} + \VV_\sn{SE}(t)
    ,
    \\
    \HH_\sn{E}(t) &= \sum_{j=1}^N \HH_{\sn{E}_j}
    ,
    \\
    \VV_\sn{SE}(t) &= \sum_{j=1}^{|\sn{E}|}\sum_{k=1}^{|\mathcal{L}(\sn{E}_j)|} \opr{S}_{jk}(t)\otimes\opr{E}_{jk}(t)
    ,
\end{align}
\label{eqn:APPGenericTdInteraction}
\end{subequations}
where we have already anticipated the requirement that the environment Hamiltonian be separable to avoid scrambling. 
Note that there are no constraints placed on the time dependence of any of the operators.
 
The most basic requirement of quantum Darwinism is the existence of a pointer observable defining a set of pointer states which are invariant under the interaction, meaning there must exist some $\opr{\mathcal{A}} = \Sop{\opr{A}}$ which commutes with $\HH_\sn{S}(t)$ and every $\opr{S}_{jk}(t)$.
Crucially, the pointer basis is a \emph{fixed} set of states, and so the pointer observable must be time-independent.
This implies that the commutators of the system operators with $\opr{A}$ must vanish for all times $t$
\begin{subequations}
\begin{align}
    \comm{\HH_\sn{S}(t)}{\opr{A}} &= 0 
    ,
    \\
    \comm{\opr{S}_{jk}(t)}{\opr{A}} &= 0 \qquad \forall j,k
    ,
\end{align}
\end{subequations}
so that the pointer states remain stable under the evolution generated by the Hamiltonian.
 
The next requirement which must be satisfied a Hamiltonian to support quantum Darwinism is that it must preserve states of singly branching form \cite{Touil22}.
In the previous App.~\ref{app:DeriveTdHamiltonians}, we derived constraints corresponding to this requirement on the operators entering a time-independent Hamiltonian by expressing the propagator as an infinite product and requiring that no factor induce intra-environmental mixing.
We take the same approach here, decomposing the propagator describing evolution under Eq.~\eqref{eqn:APPGenericTdInteraction} into an infinite product using the Wilcox expansion \cite{Blanes09}, which is essentially the Magnus expansion expanded from an exponential of a sum to a product of exponentials using the Baker-Campbell-Hausdorff formula,
\begin{align}
    \opr{U}(t) = 
        e^{-i\int_{0}^t \dd{t_1} \HH(t_1)}
        e^{-\frac{1}{2}\int_{0}^t \dd{t_1} \int_{0}^{t_1} \dd{t_0} \comm{\HH(t_1)}{\HH(t_0)}}
        \dots
\end{align}
 
Our first constraints come from examining the commutator in the argument of the second exponential.
The simplest case is when there are no free Hamiltonians, in which case this commutator expands to
\begin{widetext}
\begin{align}
    \comm{\HH(t_1)}{\HH(t_0)}
        &= \sum_{j,j',k,k'}\pqty{
            \opr{S}_{j'k'}(t_1)\opr{S}_{jk}(t_0)\otimes\comm{\opr{E}_{j'k'}(t_1)}{\opr{E}_{jk}(t_0)}
            +
            \comm{\opr{S}_{j'k'}(t_1)}{\opr{S}_{jk}(t_0)}\otimes \opr{E}_{j'k'}(t_1) \opr{E}_{jk}(t_0)
        }
        \nonumber\\
        &= \sum_{j,k,k'}\pqty{
            \opr{S}_{jk'}(t_1)\opr{S}_{jk}(t_0)\otimes\comm{\opr{E}_{jk'}(t_1)}{\opr{E}_{jk}(t_0)}
        }
            +
        \sum_{j,j',k,k'}\pqty{ 
            \comm{\opr{S}_{j'k'}(t_1)}{\opr{S}_{jk}(t_0)}\otimes \opr{E}_{j'k'}(t_1) \opr{E}_{jk}(t_0)
        }
    .
\end{align}
\end{widetext}
The first sum on the second line (where we have set $j=j'$ since environment operators acting on different degrees of freedom always commute) cannot possibly contain any intra-environmental interactions and so is compatible with the singly branching form and preserves the pointer states. 
Therefore, we conclude that just as in the time-independent case there is no requirement that the various $\opr{E}_{jk}$ commute with one another.
Similar arguments show that there are no commutativity constraints on $\HH_{\sn{E}_j}$.
 
The second sum, however, indicates that there is induced mixing between environment degrees of freedom if $\comm*{\opr{S}_{jk}(t_1)}{\opr{S}_{j'k'}(t_2)} \ne 0$ when $j\ne j'$. 
We are therefore required to impose that this commutator is identically zero for all $t_1, t_2$ and for all system operators entering $\VV_\sn{SE}(t)$.
 
If we were to include a non-trivial system Hamiltonian $\HH_\sn{S}(t)$, then upon expanding the propagator into an infinite product we would find that higher order contributions to the expansion would include integrals of time-dependent commutators of exactly the same form as in the previous appendix, e.g. $\comm{\opr{S}_{j'k'}(t_1)}{\comm{\HH_\sn{S}(t_2)}{\opr{S}_{jk}(t_3)}}$.
These commutators must vanish for all choices of evaluation times $t_1, t_2, t_3, \dots$ if intra-environmental mixing is to be avoided (provided $j\ne j'$).
Therefore we reach an analogous conclusion to the time-independent case: we must require that commuting $\comm{\HH_\sn{S}(t_2)}{\opr{S}_{jk}(t_1)}\equiv\opr{S}_{jk}'(t_1,t_2)$ arbitrarily deeply with $\HH_\sn{S}(t)$ and/or any $\opr{S}_{j'k'}(t)$ must yield zero if the corresponding set of $\opr{E}_{jk}(t)$ act on different environment subsystems. 
This must be true for any choice of evaluation time for all the operators entering the nested commutator.
Eq.~\eqref{eqn:SopStringZeroTD} in the main text expresses this statement succinctly in terms of the commutator superoperators introduced in Sec.~\ref{sec:Model}.
 
Note that as in the time-independent case, it is typically more useful to simplify these constraints to (cf. Eq.~\eqref{eqn:APPSimpleConstraints})
\begin{subequations}
\begin{align}
    \comm{\HH_\sn{S}(t_1)}{\opr{S}_{jk}(t_0)} &= 0 \qquad \forall t_1, t_0, j, k
    \\
    \comm{\opr{S}_{j'k'}(t_1)}{\opr{S}_{jk}(t_0)} &= 0 \qquad \forall t_1, t_0, j\ne j',k,k'
\end{align}
\end{subequations}
 
To conclude, we point out that the criteria we have derived which must be satisfied by a time-dependent Hamiltonian to exhibit quantum Darwinism are straightforward generalizations of those criteria we found in the time-independent case.
If the dynamics support pointer states, then those states are stable.
If further the dynamics support states of singly-branching form, then objectivity emerges eventually.
Any details of the model beyond these two facts are essentially irrelevant to the binary question of whether the asymptotic behavior exhibits quantum Darwinism or not, hence there are no fundamental differences between time-dependent and time-independent models in this respect.

\bibliography{references}
\end{document}